\documentclass[a4paper,11pt,DIV=12,abstract=true,numbers=noenddot,titlepage=false,twocolumn=false,draft=false]{scrartcl}
\pdfoutput=1

\makeatletter
\DeclareOldFontCommand{\rm}{\normalfont\rmfamily}{\mathrm}
\DeclareOldFontCommand{\sf}{\normalfont\sffamily}{\mathsf}
\DeclareOldFontCommand{\tt}{\normalfont\ttfamily}{\mathtt}
\DeclareOldFontCommand{\bf}{\normalfont\bfseries}{\mathbf}
\DeclareOldFontCommand{\it}{\normalfont\itshape}{\mathit}
\DeclareOldFontCommand{\sl}{\normalfont\slshape}{\@nomath\sl}

\usepackage[usenames,dvipsnames]{color}
  \definecolor{hgreen}{rgb}{0,.3,0}
  \definecolor{hred}{rgb}{.3,0,0}
  \definecolor{hblue}{rgb}{0,0,.3}
  \definecolor{LightGray}{gray}{0.95}
  \definecolor{gray}{gray}{0.6}
\usepackage[
	    colorlinks=true,
	    linkcolor=hblue,
	    citecolor=hgreen,
	    filecolor=hblue,
	    urlcolor=hred
	    ]{hyperref}
\usepackage{mathrsfs}
\usepackage{soul}

\usepackage[
	    colorlinks=true,
	    linkcolor=hblue,
	    citecolor=hgreen,
	    filecolor=hblue,
	    urlcolor=hred
	    ]{hyperref}
\usepackage{mathrsfs}
\usepackage[intlimits]{amsmath}
\usepackage{amssymb}
\usepackage{slashed}
\usepackage[titletoc,title]{appendix}
\usepackage[affil-it]{authblk}
\usepackage{libertine}
\usepackage[numbers,sort&compress]{natbib}
\usepackage{graphicx}
\usepackage{tensor}

\usepackage{microtype}

\setcapindent{1em}
\setkomafont{captionlabel}{\bfseries}
\setkomafont{caption}{\itshape}

\newcommand{\cO}{\mathscr{O}}
\newcommand{\Nf}{N_{\!f}}
\newcommand{\Lag}{\mathscr{L}}
\newcommand{\muew}{\mu_{\text{ew}}}

\newcommand{\q}{q}
\newcommand{\qp}{{q^{\prime}}}
\newcommand{\qpp}{{q^{\prime\prime}}}
\newcommand{\qppp}{{q^{\prime\prime\prime}}}

\definecolor{Blu}{rgb}{0.,0.,1.}

\begin{document}
\renewcommand\Authands{, }

\title{\boldmath
        Electric dipole moment constraints on CP-violating
	heavy-quark Yukawas at next-to-leading order
}
\subtitle{\flushright \vspace{-25ex} \rm DO-TH 18/22 \vspace{20ex}}

\date{\today}
\author[a,b]{Joachim Brod%
        \thanks{\texttt{joachim.brod@uc.edu}}}
\author[b,c,d]{Emmanuel Stamou%
        \thanks{\texttt{emmanuel.stamou@tu-dortmund.de}}}
	\affil[a]{{\large Department of Physics, University of Cincinnati, Cincinnati, OH 45221, USA}}
	\affil[b]{{\large Fakult\"at f\"ur Physik, TU Dortmund, D-44221 Dortmund, Germany}}
    \affil[c]{{\large Theoretical Particle Physics Laboratory (LPTP), Institute of Physics, EPFL, 1015 Lausanne, Switzerland}}
    \affil[d]{{\large Enrico Fermi Institute, University of Chicago, Chicago, IL 60637, USA}}

\maketitle

\begin{abstract}
Electric dipole moments are sensitive probes of new phases in the
Higgs Yukawa couplings. We calculate the complete two-loop QCD
anomalous dimension matrix for the mixing of CP-odd scalar and tensor
operators and apply our results for a phenomenological study of CP
violation in the bottom and charm Yukawa couplings. We find large
shifts of the induced Wilson coefficients at
next-to-leading-logarithmic order. Using the experimental bound on the
electric dipole moments of the neutron and mercury, we update the
constraints on CP-violating phases in the bottom and charm quark
Yukawas.
\end{abstract}
\setcounter{page}{1}

\section{Introduction\label{sec:introduction}}

With the discovery of the Higgs boson in 2012, the precise
determination of its couplings to all other standard model (SM)
particles became a primary goal of particle physics.  Of special
interest are quark-Yukawa couplings.  In the SM, they are real and
proportional to the quark masses; any deviation from this relation
would indicate physics beyond the SM. Such new contributions to Yukawa
couplings are often unavoidable in extensions of the SM that predict
new particles at the LHC.

Non-SM Yukawa interactions may in fact be related to the dynamics
underlying baryogenesis. For instance, new sources of CP violation
are required for electroweak baryogenesis (see
Ref.~\cite{Morrissey:2012db} for a review).  Various models of
electroweak baryogenesis require a sizeable phase in the top
Yukawa. Naively, such a phase is ruled out by its contribution to the
electric dipole moment (EDM) of the neutron. However, contributions of
the other Yukawas can cancel the contribution to EDMs without spoiling
baryogenesis~\cite{Huber:2006ri}. This motivates a detailed study of
CP-violating contributions to {\em all} Yukawa couplings.

In Ref.~\cite{Brod:2013cka}, EDM constraints on the third generation
(top, bottom, tau) Yukawa couplings were presented and compared to
bounds from Higgs production and decay at the LHC (see
Ref.~\cite{Harnik:2013aja} for a more targeted collider study for the
tau Yukawa).  Presently, anomalous bottom Yukawas are more stringently
constrained by $h\to b\bar b$ than by hadronic
EDMs~\cite{Brod:2013cka}.  EDM and collider constraints on the
electron Yukawa were subsequently studied in
Ref.~\cite{Altmannshofer:2015qra}. A more generic approach was taken
in a series of works~\cite{Chien:2015xha, Cirigliano:2016njn,
  Cirigliano:2016nyn}, in which EDM constraints on Higgs-quark and
Higgs-gluon couplings were studied in the SM effective field theory
(EFT) approach.  However, these analyses neglect logarithmically
enhanced effects that were shown to be large in
Ref.~\cite{Brod:2013cka}, and two-loop Barr--Zee contributions to the
light (up, down, and strange) Yukawas~\cite{Brod:2018lbf}.

In the present work, we address CP violation in the bottom- and charm-quark
Yukawas. The leading-logarithmic (LL) analysis for the case of
the bottom quark was performed in Ref.~\cite{Brod:2013cka}. Although
not discussed there, the residual perturbative uncertainty is
significant, exceeding the one in the non-perturbative hadronic matrix
elements.  Recent progress in lattice determinations of these matrix
elements~\cite{Bhattacharya:2015esa, Bhattacharya:2015wna,
  Bhattacharya:2015rsa, Bhattacharya:2016rrc} motivates a
next-to-leading-logarithmic (NLL) analysis in order to reduce the
perturbative error of EDM predictions for the case of beyond-the-SM CP
violation in the bottom and/or charm Yukawa. We calculate the complete
anomalous dimension matrix for the mixing of CP-odd scalar and
tensor operators up to next-to-leading (two-loop) order, and apply our
results for a phenomenological study of CP violation in the bottom
and charm Yukawa couplings.

This paper is organized as follows.  In Section~\ref{sec:eft} we
define the effective theory needed for our calculation, and present
the calculation of the renormalisation-group (RG) evolution of the
Wilson coefficients in Section~\ref{sec:rg}. We illustrate the impact
of our calculation on the constraints on the CP phases in
Section~\ref{sec:numerics}, and conclude in
Section~\ref{sec:conclusions}.  The appendices contain further details
on our work.  In Section~\ref{app:unphys} we collect the requisite
unphysical operators, in Section~\ref{app:Z} we present all relevant
renormalisation constants, and in Section~\ref{app:ADM} we present the
full anomalous-dimension matrix. We discuss effect on our results of a
change of the renormalisation scheme in
Section~\ref{app:scheme}. Section~\ref{app:expand} contains the
expansion of the resummed results.

\section{Effective theory below the weak scale}\label{sec:eft}

Our starting point is a Lagrangian in which the Higgs particle couples
to bottom or charm quarks differently than in the SM.  Such a
modification can originate from TeV-scale new physics that can be
parameterised by higher-dimensional operators, e.g., dimension-six
operators of the form
\begin{equation}\label{eq:Q:yukawa}
	H^\dagger H \bar Q_{L,3} H^\dagger d_{R,3}\,.
\end{equation}
Here, $H$ denotes the Higgs doublet in the unbroken phase of
electroweak gauge symmetry, while $Q_{L,3}$ and $d_{R,3}$ represent
the left-handed quark doublet and the right-handed down-quark field of
the third generation, respectively.

The presence of such operators induces anomalous couplings of the
Higgs particle to quarks in the electroweak broken phase. At leading
order in the electroweak interactions the above dimension-six
operators lead to a modification of the SM Yukawa interactions which,
for the bottom quark, can be conveniently parameterised as
\begin{equation}\label{eq:LagHbb}
  \Lag_{hbb} = - \frac{y^{\text{SM}}_b}{\sqrt{2}} \kappa_b \bar
  b\left( \cos\phi_b + i\gamma_5\sin\phi_b \right)b\,h\,.
\end{equation}
Here, $b$ denotes the bottom-quark field and $h$ the physical Higgs
field. Moreover, $y^{\text{SM}}_b \equiv
\frac{e}{\sqrt{2}s_w}\frac{m_b}{M_W}$ is the SM Yukawa, with $e$ the
positron charge, $s_w$ the sine of the weak mixing angle, and $m_b$
and $M_W$ the bottom-quark and $W$-boson mass, respectively.  The real
parameter $\kappa_b\geq0$ parameterises modifications to the absolute
value of the Yukawa coupling, while the phase $\phi_b \in [0,2\pi)$
parameterises CP violation and the sign of the Yukawa.  The SM
corresponds to $\kappa_b=1$ and $\phi_b=0$.  Modifications of the
charm-quark Yukawa can be parameterised in an analogous way.

The parametrisation of pseudoscalar Higgs couplings solely via the
interaction in Eq.~\eqref{eq:LagHbb} should be thought of as the
dimension-four part of the so-called Higgs Effective Field
Theory~\cite{Feruglio:1992wf} or the electroweak chiral
Lagrangian\cite{Buchalla:2013rka} in {\it unitarity gauge} for the
electroweak sector. In this sense, the SM extended by a pseudoscalar
Higgs coupling is not a renormalizable theory. However, as we consider
in this work only loops induced by the strong coupling, we do not
encounter additional divergences since the operators of the
form~\eqref{eq:Q:yukawa} mix only into themselves under
QCD~\cite{Alonso:2013hga}. This picture would change if one considered
higher-order electroweak corrections.

\begin{figure}[t]
        \centering
        \includegraphics{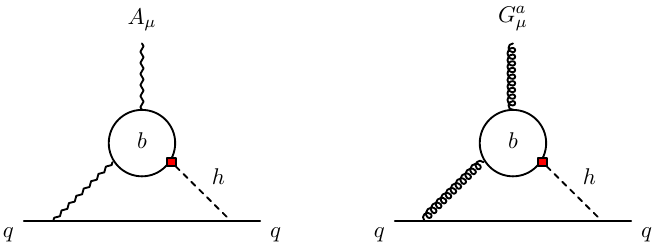}
        \caption{Photonic and gluonic ``Barr--Zee'' diagrams with
          modified bottom-Yukawa coupling that induce an EDM of the
          light quark $q$. See text for details.
	\label{fig:barrzee}}
\end{figure}

The basic idea underlying this work is to calculate the effect of
CP-odd phases in the Higgs Yukawa couplings of the bottom and charm
quark on hadronic EDMs.  EDMs receive contributions from partonic
CP-violating electric and chromoelectric dipole operators, with
coefficients $d_q$ and $\tilde d_q$. They are traditionally defined
via the effective Lagrangian valid at hadronic energies $\mu \simeq
2\,$GeV~\cite{Engel:2013lsa},
\begin{equation} \label{eq:LeffN}
\Lag_{\rm eff} = - d_q \, \frac{i}{2} \, \bar q \sigma^{\mu\nu}
\gamma_5 q \, F_{\mu\nu} - \tilde d_q \, \frac{ig_s}{2} \, \bar q
\sigma^{\mu\nu} T^a \gamma_5 q \, G_{\mu\nu}^a + \frac{1}{3} w f^{abc}
\, G_{\mu \sigma}^a G_{\nu}^{b, \sigma} \widetilde G^{c, \mu \nu}\,,
\end{equation}
with $\sigma^{\mu\nu} = \tfrac{i}{2} [\gamma^\mu, \gamma^\nu]$, as
well as $T^a$ the fundamental generators of SU$(n_c)$ with
Tr$[T^a,T^b]= \delta^{ab}/2$ and $n_c=3$ the number of colors. This
Lagrangian also includes the purely gluonic Weinberg
operator~\cite{Weinberg:1989dx}. Its contributions are subdominant
because of its small nuclear matrix elements~\cite{Pospelov:2005pr,
  Engel:2013lsa}, but are kept for completeness.

CP-violating Yukawa couplings contribute to $d_q$ and $\tilde{d}_q$
via Barr--Zee-type diagrams~\cite{Barr:1990vd} with heavy-quark loops,
see Fig.~\ref{fig:barrzee}. The Weinberg operator is induced via a
finite threshold correction~\cite{Braaten:1990gq, Braaten:1990zt}. In
the case of a CP-violating bottom Yukawa, expanding the loop functions
for small bottom mass and matching directly onto the Lagrangian of
Eq.~\eqref{eq:LeffN}, we find
\begin{equation} \label{eq:dsw}
\begin{split}
d_q & \simeq -12 e Q_q Q_b^2 \, \frac{\alpha}{(4\pi)^3} \sqrt{2}
G_F m_q \, \kappa_b \sin\phi_b \, x_b \left (
\log^2 x_b + \frac{\pi^2}{3} \right ) \,,\\[0.5em]
\tilde d_q & \simeq 2\frac{\alpha_s}{(4\pi)^3} \sqrt{2} G_F m_q
\, \kappa_b \sin\phi_b \, x_b \left ( \log^2
x_b + \frac{\pi^2}{3} \right ) \,,\\[0.5em]
w & \simeq - g_s \frac{\alpha_s}{(4\pi)^3} \sqrt{2} G_F
x_b \, \kappa_b^2 \cos\phi_b \, \sin\phi_b \,
\bigg(\frac{3}{2} + \log x_b \bigg) \,,
\end{split}
\end{equation}
up to higher orders in $x_b \equiv m_b^2/M_h^2$.

However, as already noted in Ref.~\cite{Brod:2013cka}, such a naive
evaluation of the gluonic diagram leads to an uncertainty of a factor
of order five.  The uncertainty is related to the ambiguity in
choosing the proper value of the strong coupling $\alpha_s(\mu)$;
namely, at which dynamical scale should it be evaluated -- the weak
scale, the bottom-quark mass, or the hadronic scale?  This scale
dependence is related to logarithms of the large scale ratios and can
be reduced by resummation of the large logarithms, which is easiest
performed in an effective theory (EFT) framework.  The LL series then
reproduces the quadratic logarithm in Eq.~\eqref{eq:dsw}, while also
resumming all higher-order terms. The uncertainties after the LL
resummation are still large, at the order of two at the Wilson
coefficient level. Hence, in this work we extend the LL analysis of
Ref.~\cite{Brod:2013cka} to NLL and discuss the remaining theory
uncertainty via the residual scale and scheme dependence in detail. In
addition, we consider also modifications of the charm-quark Yukawa.

To construct the effective Lagrangian originating from anomalous
flavor-conserving, CP-violating Higgs couplings to quarks we
integrate out the heavy degrees of freedom of the SM (the top quark,
the weak gauge bosons, and the Higgs) and match onto an effective
five-flavor Lagrangian.  EDMs are then induced by non-renormalizable
operators that are CP odd.  The corresponding Lagrangian reads
\begin{multline} \label{eq:Leff}
\Lag_\text{eff} =
  - \sqrt{2} G_F\, \biggl(
	\sum_{q\neq q'}
	  \bigg[ \sum_{i=1,2} C_i^{qq'}  O_i^{qq'} + \frac{1}{2} \sum_{i=3,4} C_i^{qq'}  O_i^{qq'} \bigg]
 + \sum_q  \sum_{i=1,\ldots,4} C_i^q   O_i^q  + C_w  O_w \bigg ) \,,
\end{multline}
where the sums run over all quarks with masses below the weak scale
($q,q' = u,d,s,c,b$).
The linearly independent operators are
\begin{align}
\label{eq:op:qq':1:2}
O_1^{qq'} & = (\bar q q) \, (\bar q' \, i \gamma_5 q')\,,
&
O_2^{qq'} & = (\bar q \, T^a  q) \, (\bar q' \, i \gamma_5 T^a q') \,,\\[0.5em]
\label{eq:op:qq':3:4}
O_3^{qq'} & = \frac{1}{2}\epsilon^{\mu\nu\rho\sigma} (\bar q \sigma_{\mu\nu} q) \, (\bar q' \, \sigma_{\rho\sigma} q') \,,
&
O_4^{qq'} & = \frac{1}{2}\epsilon^{\mu\nu\rho\sigma}(\bar q \sigma_{\mu\nu} T^a q) \, (\bar q' \sigma_{\rho\sigma} T^a q')\,,\\[0.5em]
\label{eq:op:q}
O_1^q & = (\bar q q) \, (\bar q \, i \gamma_5 q) \,,
&
O_2^q & = \frac{1}{2}\epsilon^{\mu\nu\rho\sigma}(\bar q \sigma_{\mu\nu} q) \, (\bar q \, \sigma_{\rho\sigma} q) \,,\\[0.5em]
O_3^q & = \frac{eQ_q}{2} \frac{m_q}{g_s^2} \, \bar q \sigma^{\mu \nu} q \, \tilde{F}_{\mu \nu} \,,
&
O_4^q & =  -\frac{1}{2} \, \frac{m_q}{g_s} \, \bar q \sigma^{\mu \nu} T^a q \, \tilde{G}^a_{\mu \nu} \,,
	\label{eq:dipoles}
\end{align}
and
\begin{equation} \label{eq:op:weinberg}
\begin{split}
O_w & = -\frac{1}{3\,g_s} f^{abc} \, G_{\mu \sigma}^a G_{\nu}^{b, \sigma} \widetilde G^{c, \mu \nu} \,.
\end{split}
\end{equation}
The basis of all CP-odd operators is closed under the
renormalisation group flow of both QCD and QED as they both conserve
CP.

In these equations, the $\gamma_5$ matrix is defined by
\begin{equation}\label{eq:gamma5:def}
  \gamma_5 = \frac{i}{4!} \epsilon_{\mu\nu\rho\sigma} \gamma^\mu
  \gamma^\nu \gamma^\rho \gamma^\sigma\,,
\end{equation}
where $\epsilon^{\mu\nu\rho\sigma}$ is the totally antisymmetric
Levi-Civita tensor in four space-time dimensions with
$\epsilon_{0123}=-\epsilon^{0123}=1$, and we use the notation
$\widetilde G^{a, \mu \nu} = \tfrac{1}{2} \epsilon^{\mu\nu\rho\sigma}
G_{\rho\sigma}^a$.  We have included the factor $1/2$ in the
contributions of the $O_{3(4)}^{qq'}$ operators to the effective
Lagrangian to account for the double counting implied by the sums in
Eq.~\eqref{eq:Leff}. The non-standard sign convention for $O_3^q$ is
related to our definition of the covariant derivative,
Eq.~\eqref{eq:codev}.

Note that whenever possible we defined the operators directly in terms
of the Levi-Civita tensors instead of $\gamma_5$ matrices. This
reduces the number of $\gamma$ matrices entering the computation of
the anomalous dimensions.  In $d=4$ dimensions our definition is
equivalent to the usual one because of the relation
\begin{equation}
  \sigma^{\mu\nu} \gamma_5 \overset{[d=4]}{=} -\frac{i}{2} \epsilon^{\mu\nu\rho\sigma}
  \sigma_{\rho\sigma}\,,
\end{equation}
valid in $d=4$. While such evanescent differences in the definition of
the operators do not affect the leading-order anomalous dimensions,
they do affect the next-to-leading-order results that we compute in
this work. For further details regarding our treatment of $\gamma_5$
see section~\ref{sec:RG}.

\section{Renormalisation group evolution}\label{sec:rg}

Our goal is the summation of all leading and next-to-leading
logarithms via RG improved perturbation theory in the five-, four-,
and three-flavor effective theory.  The calculation proceeds in
several steps.  First, we integrate out the Higgs and weak gauge
bosons together with the top quark at the electroweak scale, $\mu \sim
M_h = 125.10\,$GeV. After the RG flow, the heavy (bottom and charm)
quarks are integrated out at their respective masses, $m_b(m_b) =
4.18\,$GeV and $m_c(m_c) = 1.27\,$GeV (all numerical input values are
taken from Ref.~\cite{Zyla:2020zbs}). We then match to an effective
three-flavor theory where only the light-quark operators are
present. Finally, we evaluate the Wilson coefficients in the
three-flavor theory at $\mu_\text{str} = 2\,$GeV where the hadronic
matrix elements of the electric dipole operators are given by the
lattice calculations. The RG evolution between the different matching
scales is computed using the appropriate anomalous dimensions,
following the general formalism outlined in
Ref.~\cite{Buchalla:1995vs}.

The actual calculation was performed with self-written
\texttt{FORM}~\cite{Vermaseren:2000nd} routines, implementing the
two-loop recursion presented in Refs.~\cite{Davydychev:1992mt,
  Bobeth:1999mk}. The amplitudes were generated using
\texttt{QGRAF}~\cite{Nogueira:1991ex}.

\subsection{Initial conditions at the weak scale}

\begin{figure}[t]
	\centering
	\includegraphics{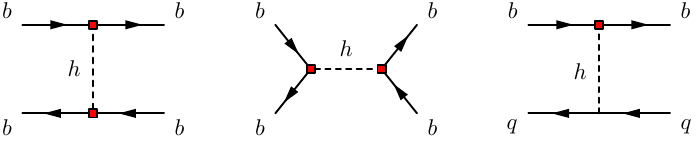}
	\caption{Sample tree-level Feynman diagrams contributing to
          the calculation of the initial conditions of the RG
          evolution at the electroweak scale for the case of modified
          bottom-quark Yukawas (indicated by the red square). Light
          quarks are denoted by the label $q=u,d,s$.
	\label{fig:matching0loop}}
\end{figure}

\begin{figure}[t]
	\centering
	\includegraphics{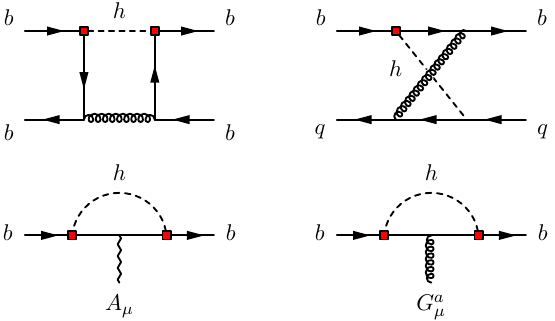}
	\caption{Sample one-loop Feynman diagrams contributing to the
          calculation of the initial conditions of the RG evolution at
          the electroweak scale for the case of modified bottom-quark
          Yukawas (indicated by the red square). Light quarks are
          denoted by the label $q=u,d,s$.
	\label{fig:matching1loop}}
\end{figure}

We augment the SM with flavor-conserving anomalous Higgs Yukawas
parameterised as in Eq.~\eqref{eq:LagHbb} and at a scale $\muew\approx
M_h$ integrate out the heavy degrees of freedom of the SM. Up to
quadratic order in the strong coupling constant, we find the following
non-zero initial conditions for the Wilson coefficients at scale
$\muew$:
\begin{align}
C_1^{\q\qp}(\muew) &= - \kappa_\q \kappa_\qp \cos\phi_\q \sin\phi_\qp\frac{m_q m_{q'}}{M_h^2} + {\cal O} (\alpha_s^2)\,,\label{eq:init:C1qqp}\\
C_4^{\q\qp}(\muew) &=  \frac{\alpha_s}{4\pi}
                        \left(\frac{3}{2}+ \log\frac{\muew^2}{M_h^2}   \right)\frac{m_q m_{q'}}{M_h^2}
			\kappa_\q \kappa_\qp (\cos\phi_\q \sin\phi_\qp + \sin\phi_\q \cos\phi_\qp)
		       + {\cal O} (\alpha_s^2)\,,\\
C_1^{\q}(\muew)    &= - \left(
1 + \frac{\alpha_s}{4\pi}\left( \frac{9}{2}+3\log\frac{\muew^2}{M_h^2} \right)\right)\frac{m_q^2}{M_h^2} \kappa_\q^2 \cos\phi_q\sin \phi_\q
		       + {\cal O} (\alpha_s^2)\,,\\
C_2^{\q}(\muew)    &=  \frac{\alpha_s}{4\pi}\left( \frac{1}{8}+\frac{1}{12}\log\frac{\muew^2}{M_h^2} \right)\frac{m_q^2}{M_h^2} \kappa_\q^2 \cos\phi_q\sin \phi_\q
		       + {\cal O} (\alpha_s^2)\,,\\
C_3^{\q}(\muew)    &= -\frac{\alpha_s}{4\pi}\left( 3+2\log\frac{\muew^2}{M_h^2} \right)\frac{m_q^2}{M_h^2} \kappa_\q^2 \cos\phi_q\sin \phi_\q
		       + {\cal O} (\alpha_s^2)\,,\\
C_4^{\q}(\muew)    &= -\frac{\alpha_s}{4\pi}\left( 3+2\log\frac{\muew^2}{M_h^2} \right)\frac{m_q^2}{M_h^2} \kappa_\q^2 \cos\phi_q\sin \phi_\q
		       + {\cal O} (\alpha_s^2)\,.\label{eq:init:C4q}
\end{align}
We see that if a single CP-violating Yukawa is switched on, e.g.,
the $q$-Yukawa with $q=b,c$, then at tree-level only the operators
$O_1^q$ and $O_1^{q'q}$ are induced by the anomalous Higgs coupling to
$q$, i.e., to bottom or charm quarks, see
Fig.~\ref{fig:matching0loop}.  At one loop, also the operators
$O_{2\ldots 4}^q$ and $O_4^{qq'}$ receive non-zero initial conditions
(Fig.~\ref{fig:matching1loop}).

The quark masses in the expressions above are understood to be
evaluated at the matching scale $\mu=\muew$. Note that the one-loop
expressions depend on the definition of evanescent operators; our
choice is given in App.~\ref{app:unphys}. Our predictions of physical
observables does not depend on the renormalisation scheme to ${\cal O}
(\alpha_s)$; we discuss the residual scheme dependence in
Section~\ref{sec:numerics}.

\subsection{Calculation of the anomalous dimensions}\label{sec:RG}

The RG evolution below the weak scale in the presence of the heavy
quarks can be calculated from the mixing of the operators in
Eq.~\eqref{eq:Leff}.  The one-loop results are fully known (see, e.g.,
Ref.~\cite{Hisano:2012cc}). The two-loop mixing of the dipole
operators has been presented in Refs.~\cite{Misiak:1994zw,
  Gorbahn:2005sa}. As a cross check we have reproduced all these
results in the literature (see App.~\ref{app:scheme} for details). The
two-loop mixing of the four-fermion operators among themselves and
into the dipole operators is presented here for the first time.

\begin{figure}[t]
	\centering
	\includegraphics{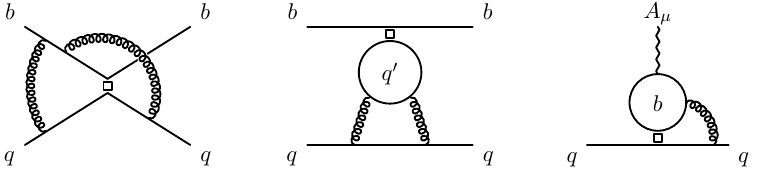}
	\caption{Sample two-loop Feynman diagrams whose divergent
          parts contribute to the calculation of the two-loop
          anomalous dimensions involving bottom quarks. The empty
          squares symbolize the insertion of an effective
          operator. The diagrams of the kind shown in the middle
          panel, involving three different quark flavors, vanish due
          to the odd number of Dirac matrices in the trace. Diagrams
          of the kind shown in the right panel contain traces of Dirac
          matrices with $\gamma_5$, which are treated in the Larin
          scheme.
	\label{fig:running2loop}}
\end{figure}

We calculate the anomalous dimensions by extracting the divergent
parts of the insertions of all operators into appropriate Greens
functions (see Fig.~\ref{fig:running2loop} for sample diagrams
involving the bottom quark).  We use dimensional regularisation,
working in $d=4-2\epsilon$ space-time dimensions, and employ the
method of infrared rearrangement (IRA) described in
Ref.~\cite{Chetyrkin:1997fm} to disentangle infrared from ultraviolet
poles. For the whole computation we work in generic $R_{\xi_g}$ gauge
for SU$(3)_c$, which provides further checks for the correctness of
the computation.

The appearance of $\gamma_5$ in closed fermion loops requires special
care, since the use of a naively anticommuting $\gamma_5$, with
$\{\gamma^{\mu},\gamma_5\} = 0$ for all $\mu$ (NDR scheme) is
algebraically inconsistent if traces with $\gamma_5$ have to be
evaluated~\cite{Collins:1984xc}. Since such traces appear in our
calculation we use Larin's prescription~\cite{Larin:1993tq}
throughout: The Levi-Civita tensor appearing either in the definition
of $\gamma_5$ or in the effective operators is taken outside the
momentum integrals and acts as a projector on the four-dimensional
physical subspace {\em after} the momentum integration has been
performed. The momentum integration, including the Dirac algebra, is
then performed in $d$ space-time dimensions. The contraction of the
Levi-Civita tensor with the results of the momentum integrals
generates several evanescent operators that affect the two-loop
anomalous dimensions in the usual way; they are listed in
App.~\ref{app:unphys}. No more than one Levi-Civita tensor appears in
any diagram of the computation, so there is no need to contract
multiple tensors.

The RG flow of the Wilson coefficients is governed by the RG equation
\begin{equation}
	\frac{d {C^t}}{d\log\mu} =  C^t \gamma
	\quad\text{with}\quad \gamma=\frac{\alpha_s}{4\pi}\gamma^{(0)} +
	\left(\frac{\alpha_s}{4\pi}\right)^2\gamma^{(1)} +\ldots\,,
\end{equation}
where the superscript $t$ denotes transposition and $\gamma^{(0)}$,
$\gamma^{(1)}$ are the one- and two-loop anomalous-dimension matrices,
respectively.  For our NLL analysis we require the two-loop mixing of
the four-quark operators among themselves and into the dipole
operators.

Below we collectively use the subscripts $qq'$ and $q$ in the
$\gamma$'s to indicate subblocks of the full anomalous-dimension
matrix.  The ordering in each case is
\begin{equation}
C_{qq'} = \{C_1^{qq'},\, C_2^{qq'},\, C_1^{q'q},\, C_2^{q'q},\,C_3^{qq'},\, C_4^{qq'}\}\,,
\quad \text{and}\quad
C_q = \{C_1^{q},\,C_2^{q},\,C_3^{q},\,C_4^{q}\}\,.
\end{equation}
With the definition of evanescent operators given in App.~\ref{app:unphys},
we find the mixing among four-fermion operators with two different flavors to
be
\begin{align}
\gamma^{(1)}_{\q\qp\to\q\qp} &=
\left(
\begin{matrix}
-\frac{16 N_f}{3}-36 & -\frac{5}{2} & -\frac{28}{3} \\[0.5em]
 -\frac{5}{9} & \frac{1045}{12}-\frac{16 N_f}{3} & -\frac{35}{9} \\[0.5em]
 -\frac{28}{3} & -\frac{35}{2} & -\frac{16 N_f}{3}-36 \\[0.5em]
 -\frac{35}{9} & \frac{1}{4} & -\frac{5}{9} \\[0.5em]
 0 & \frac{16 N_f}{3}-24 & 0 \\[0.5em]
 \frac{32 N_f}{27}-\frac{496}{3} & \frac{20 N_f}{9}-190 & \frac{32 N_f}{27}-\frac{496}{3}
\end{matrix}
\right.\nonumber\\[1em]
	\label{eq:ADM:NLO:qqp:qqp}
& \hspace{5em}
\left.
\begin{matrix}
 -\frac{35}{2} & 0 & \frac{2 N_f}{9}-37 \\[0.5em]
 \frac{1}{4} & \frac{4 N_f}{81}+\frac{10}{9} & \frac{5 N_f}{54}-\frac{95}{12} \\[0.5em]
 -\frac{5}{2} & 0 & \frac{2 N_f}{9}-37 \\[0.5em]
 \frac{1045}{12}-\frac{16 N_f}{3} & \frac{4 N_f}{81}+\frac{10}{9} & \frac{5 N_f}{54}-\frac{95}{12} \\[0.5em]
 \frac{16 N_f}{3}-24 & \frac{1288}{9}-\frac{208 N_f}{27} & -\frac{100}{3} \\[0.5em]
 \frac{20 N_f}{9}-190 & -\frac{200}{27} & -\frac{100 N_f}{27}-\frac{340}{9}
\end{matrix}
\right)\,.\\
\intertext{We find for their mixing into four-quark operators with one quark flavor and into dipoles}
\label{eq:ADM:NLO:qqp:q}
\gamma^{(1)}_{\q\qp\to\q} &=
\begin{pmatrix}
 0 & 0 & 0 & -12\frac{m_\qp}{m_\q} \\[0.5em]
 0 & 0 & -16\frac{Q_\qp}{Q_\q}\frac{m_\qp}{m_\q} & -5\frac{m_\qp}{m_\q} \\[0.5em]
 0 & 0 & 0 & -4\frac{m_\qp}{m_\q} \\[0.5em]
 0 & 0 & -\frac{16}{3}\frac{Q_\qp}{Q_\q}\frac{m_\qp}{m_\q} & -\frac{5}{3}\frac{m_\qp}{m_\q} \\[0.5em]
 0 & 0 & -448\frac{Q_\qp}{Q_\q}\frac{m_\qp}{m_\q} & 0 \\[0.5em]
 0 & 0 & \frac{128}{3} \frac{m_\qp}{m_\q} & - 92 \frac{m_\qp}{m_\q}
\end{pmatrix}\,.
\intertext{For the anomalous dimensions in the $q\to q$ sector we find}
\gamma^{(1)}_{\q\to\q} &=
\begin{pmatrix}
 65-6 \Nf & \frac{\Nf}{54}-\frac{19}{12} & \frac{848}{27} & \frac{1235}{27} \\[0.5em]
 60-\frac{40 \Nf}{9} & \frac{403}{3}-\frac{226 \Nf}{27} & -\frac{6976}{9} & -\frac{820}{9} \\[0.5em]
 0 & 0 & \frac{460 \Nf}{27}+\frac{100}{9} & 0 \\[0.5em]
 0 & 0 & 112-\frac{128 \Nf}{27} & \frac{449 \Nf}{27}+\frac{577}{9}
\end{pmatrix}\,.
\end{align}
The two-loop mixing among the dipole operators $O_3^q$ and $O_4^q$ has
been calculated before; the relation between our results and the
literature is discussed in App.~\ref{app:scheme}. All remaining
results are new.

The two-loop ADM that involve the Weinberg operator and do not
necessarily vanish are $\gamma^{(1)}_{W\to\q}$,
$\gamma^{(1)}_{W\to\q\qp}$, and $\gamma^{(1)}_{W\to W}$. They are not
needed in our analysis, since the Weinberg operator contributes only
via a finite threshold correction, see Section~\ref{sec:match} (the
two-loop and three-loop self mixings of the Weinberg operator for the case of $\Nf=0$
have recently been published for pure Yang-Mills theory in
Ref.~\cite{deVries:2019nsu}). All remaining subblocks are zero, i.e.,
\begin{equation}
\gamma^{(1)}_{\q\to\qp} =
\gamma^{(1)}_{\q\to\q\qp} =
\gamma^{(1)}_{\q\to\qp\qpp} =
\gamma^{(1)}_{\q\qp\to\qpp} =
\gamma^{(1)}_{\q\qp\to\qp\qpp} =
\gamma^{(1)}_{\q\qp\to\qpp\qppp} =
\gamma^{(1)}_{\q\to W} =
\gamma^{(1)}_{\q\qp\to W} =
\boldsymbol{0}\,.
\end{equation}
The appearance of quark-mass ratios in these matrices is related to
the explicit factors of quark masses in Eq.~\eqref{eq:dipoles}. These
mass ratios are scale- and scheme independent to the order we are
working. They could, in principle, be avoided altogether by defining
several dipole operators with the same field content, but different
quark mass factors.

For completeness, we collect the one-loop anomalous dimensions in
App.~\ref{app:ADM}.

\subsection{Matching at the heavy-quark thresholds}\label{sec:match}

If $m_q \ll m_{q'}$, the dipole operators with external $q$-quark
lines receive matching corrections at the $q'$-quark threshold. (In
practice, $q'=b$ or $q'=c$.) We write the effective Lagrangian of the
theory in which all heavy $q'$ quarks have been integrated out as
\begin{equation} \label{eq:Leff:n-1}
\Lag_\text{eff} =
  - \sqrt{2} G_F \bigg( \sum_q  \sum_{i=3,4} C_i^q   O_i^q + C_w O_w + \ldots \bigg) \,.
\end{equation}
Here, $q=u,d,s$ denotes one of the light quarks. CP-violating
four-fermion operators involving only light quarks are present in
principle and denoted by the ellipsis; however, their Wilson
coefficients are suppressed by a power of $m_q/m_{q'}$.

Let us call the theory with and without the $q'$ quark the
$f$-flavor and $(f-1)$-flavor theories, respectively.
At the threshold scale $\mu_f$ the amplitudes in both theories
must match each other; we write the equality between
two general amplitudes in these theories as
\begin{equation}\label{eq:matching:condition}
\sum_i C_{i,f} (\mu_f) \langle O_{i} \rangle_f (\mu_f) =
\sum_i C_{i,f-1} (\mu_f) \langle O_i \rangle_{f-1} (\mu_f) \,.
\end{equation}
Here, angle brackets denote appropriate matrix elements. We expand the
Wilson coefficients and matrix elements in the strong coupling of the
$f$-flavor theory:
\begin{align}
C_{i,f} (\mu_f) & = C_{i,f}^{(0)} + \frac{\alpha_s(\mu_f)}{4\pi}
C_{i,f}^{(1)} + \ldots \,,\\
\langle O_{i} \rangle_f (\mu_f) & = \sum_j \bigg( \delta_{ij} +
\frac{\alpha_s(\mu_f)}{4\pi} r_{ij,f}^{(1)} + \ldots \bigg) \langle
O_{j} \rangle_f^{(0)} (\mu_f) \,,
\end{align}
where the ellipses denote higher orders in the strong coupling
constant. We expressed the higher-order matrix elements in terms of
tree-level matrix elements, denoted by the superscript ``$(0)$'', via
the coefficients $r_{ij}$.

There are various subtleties to keep in mind when calculating the
threshold corrections. The strong coupling constant itself receives a
non-vanishing threshold correction at one loop,
\begin{equation}
\alpha_s^{(f)}(\mu_f) = \alpha_s^{(f-1)}(\mu_f) \bigg( 1 +
\frac{\alpha_s^{(f-1)}(\mu_f)}{4\pi} \frac{2}{3}
\log\frac{\mu_f^2}{m_{q'}^2} + \ldots \bigg) \,.
\end{equation}
Similarly, the gluon field renormalisation receives a non-zero
threshold correction. Finally, the anomalous dimensions of the dipole
operators depend explicitly on the number of active quark flavors.
The quark masses, on the other hand, are affected only at the two-loop
level (see, e.g., Ref.~\cite{Chetyrkin:1997un}).

\begin{figure}[t]
	\centering
	\includegraphics{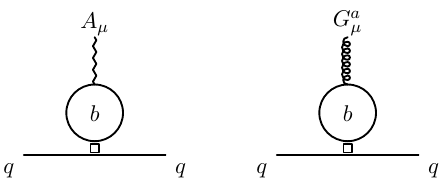}
	\caption{Sample Feynman diagrams whose finite parts contribute
          to the calculation of the matching corrections at the
          respective heavy-quark thresholds (shown here for the case
          of the bottom quark).
	\label{fig:matchingflavors}}
\end{figure}

By explicit calculation of various one-loop diagrams (see
Fig.~\ref{fig:matchingflavors}), we find, evaluating
Eq.~\eqref{eq:matching:condition}, the following non-zero
threshold corrections
\begin{align}
C_{3,f-1}^{q,(1)} & = C_{3,f}^{q,(1)} + \bigg( 24 C_{3,f}^{qq',(0)} (\mu_f)
\frac{m_{q'}}{m_q} \frac{Q_{q'}}{Q_q} - \frac{2}{3} C_{3,f}^{q,(0)}
(\mu_f) \bigg) \log\frac{\mu_f^2}{m_{q'}^2} \,, \\
C_{4,f-1}^{q,(1)} & = C_{4,f}^{q,(1)} + \bigg( 4 C_{4,f}^{qq',(0)} (\mu_f)
\frac{m_{q'}}{m_q} - \frac{2}{3} C_{4,f}^{q,(0)} (\mu_f) \bigg)
\log\frac{\mu_f^2}{m_{q'}^2} \,, \\
C_{w,f-1}^{(1)} & = C_{w,f}^{(1)} - \frac{4}{3} C^{(0)}_{w,f}(\mu_f)\log \frac{\mu_f^2}{m_{q'}^2}
 - \frac{1}{2} C_{4,f}^{q',(0)}(\mu_f) \,,
\end{align}
as a power series in the strong coupling in the $(f-1)$-flavor theory.

\section{Numerics}\label{sec:numerics}

In the last three sections we presented all the ingredients needed to
consistently perform the resummation of large logarithms appearing in
hadronic Barr--Zee-type diagrams from flavor-conserving CP-violating
Higgs Yukawas at NLL: the next-to-leading-order (NLO)
(one-loop) matching at the electroweak scale, the NLO (two-loop)
anomalous-dimension matrix, and the (one-loop) threshold corrections
for the Wilson coefficients at the heavy-quark thresholds.  In this
section, we implement the NLL evolution numerically and discuss its
impact on a set of hadronic EDMs. We first present values for the
partonic Wilson coefficients in dependence of the CP-violating phase
and discuss the theoretical uncertainty in detail. Then we give bounds
on the phases using experimental input.

\subsection{Wilson coefficients}\label{sec:wilson}

To compute the effect of modified Yukawa couplings on hadronic EDMs,
we need the values of the induced Wilson coefficients of the electric
dipole, chromoelectric dipole, and Weinberg operators at the scale
$\mu_{\text{str}} = 2\,$GeV where the matrix elements of these
operators are evaluated.  We consider two cases: first, we only turn
on a CP-violating bottom Yukawa and second, only a CP-violating charm
Yukawa.

The dependence on the matching scales $\muew$ and $\mu_{b(c)}$ of the
dipole Wilson coefficients evaluated at scale $\mu_{\text{str}} =
2\,$GeV cancels at NLO in the expansion in $\alpha_s$.\footnote{In our
  case, the Weinberg operator does not contribute at LL. Hence, there
  is no corresponding cancellation to the order we calculated.}
However, the RG evolution induces a residual dependence on these
scales. This dependence is formally of higher order in $\alpha_s$ and
can be used to assess the remaining theoretical uncertainty of the
prediction.
\begin{figure}[t!]
	\includegraphics{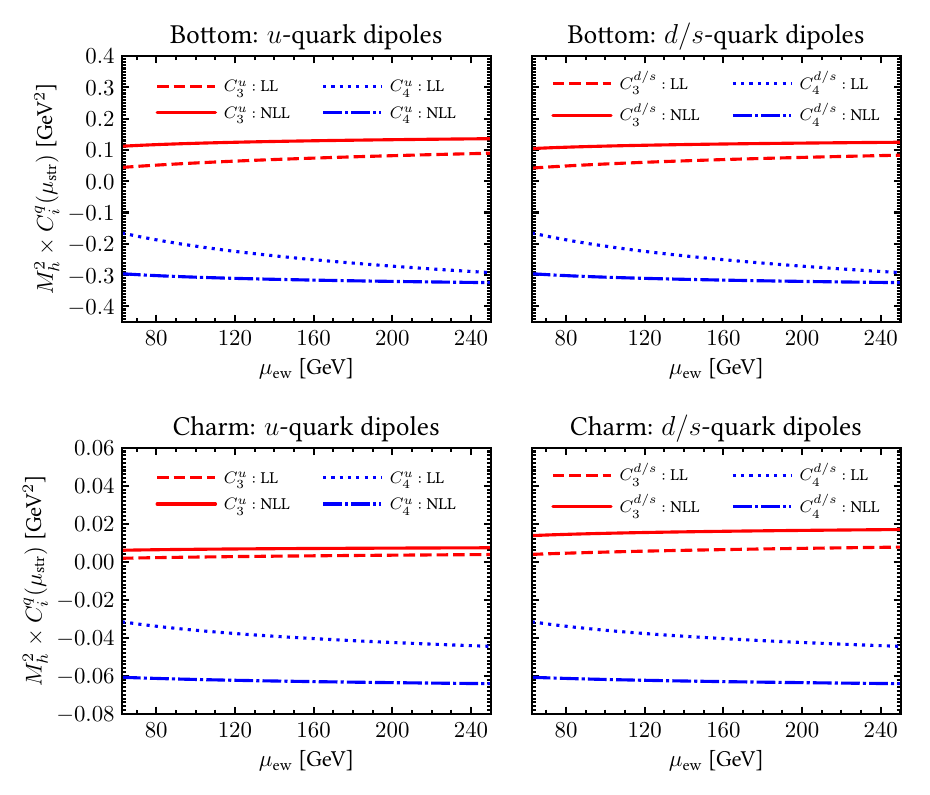}
	\caption{Residual dependence of the dipole Wilson coefficients
          on the electroweak matching scale.  The upper two panels
          show the effect of a modified bottom Yukawa, the lower two
          panels the effect of a modified charm Yukawa.  The left
          plots correspond to operators with external up quarks, while
          the right plot show operators with down- / strange-quark
          external legs.  In all cases the Wilson coefficients are
          evaluated in the three-flavor theory at the hadronic scale
          $\mu_{\text{str}}=2$\,GeV, for $\phi_{b(c)}=\pi/2$ and
          $\kappa_{b(c)}=1$.
	\label{fig:muhdependence}
	  }
\end{figure}
In Fig.~\ref{fig:muhdependence} we show the value of the dipole Wilson
coefficients, evaluated at $2$~GeV, as a function of the electroweak
matching scale, $\muew$, varied within $M_h/2 \leq \muew \leq
2M_h$. The Wilson coefficients contain terms proportional to either
$\kappa_{b(c)}\sin\phi_{b(c)}/M_h^2$ or
$\kappa^2_{b(c)}\sin\phi_{b(c)}\cos\phi_{b(c)}/M_h^2$.  For the dipole
operators, the latter terms are subleading.  For purposes of
illustration we thus choose to plot the case $\phi_{b(c)}=\pi/2$,
setting $\kappa_{b(c)}=1$ and factoring out the global Higgs-mass
dependence.  Focusing first on the Wilson coefficients $C_3^q$ of the
electric dipole operators (red lines), we see that the scale
dependence is both weak and barely reduced by going from LL to NLL.
For the coefficients $C_4^q$ of the chromoelectric dipole operator
(blue lines) the scale dependence $\delta C_4^q/C_4^q$ is reduced by
factor of approximately five. In all cases, however, the shift from
the LL to the NLL results is substantial and larger than indicated by
the residual scale dependence of both the LL and NLL result.

It is interesting to note that the RG evolution, in the case of a
modified charm Yukawa, leads to a small Wilson coefficient of the
electric dipole operator, such that the contribution to the hadronic
EDMs at low scales is completely dominated by the chromoelectric
dipole operator.

\begin{figure}[t!]
	\includegraphics{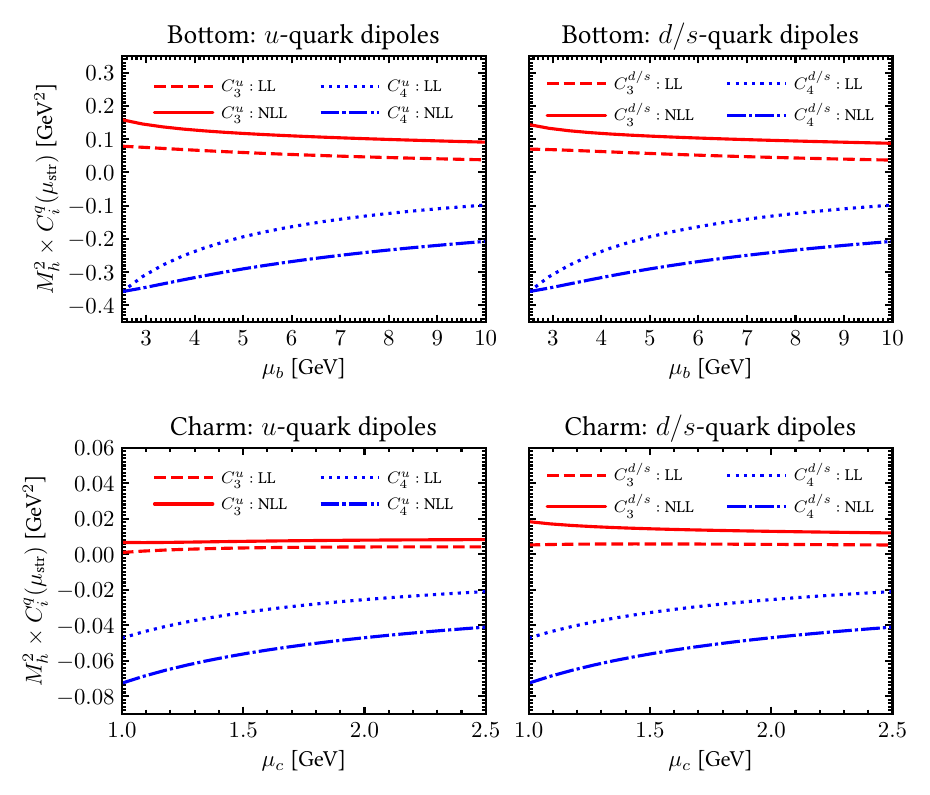}
	\caption{The same as Fig.~\ref{fig:muhdependence}, but the
          dependence on the matching scale at the bottom- and
          charm-quark thresholds is shown.
	\label{fig:mubcdependence}}
\end{figure}

Next, we consider the residual dependence on the matching scale at
which the heavy quark is integrated out. Similarly to
Fig.~\ref{fig:muhdependence}, in Fig.~\ref{fig:mubcdependence} we show
the Wilson coefficients at $2$~GeV, this time varying the heavy-quark
matching scales within $2.5\,\text{GeV} \leq \mu_b \leq
10\,\text{GeV}$ and $1\,\text{GeV} \leq \mu_c \leq
2.5\,\text{GeV}$. While the scale dependence of the electric dipole
coefficients $C_3^q$ is still mild (red lines), the chromoelectric
coefficients $C_4^q$ (blue lines) show a large residual scale
dependence that is barely reduced in going from LL to NLL. Again, all
Wilson coefficients show a large shift when including the NLL
corrections.

While the large scale dependence at NLL can be partly understood by
the appearance of many new, non-zero electroweak initial conditions at
NLO (cf.~Eqs.~\eqref{eq:init:C1qqp}--\eqref{eq:init:C4q}), the large
shift is mainly due to the large numerical values of the entries in
the NLO anomalous-dimension matrix (see, e.g.,
Eq.~\eqref{eq:ADM:NLO:qqp:q}).  This suspicion is borne out by
expanding the result of the RG evolution about the bottom-quark
matching scale, illustrating the size of the higher-order corrections;
see App.~\ref{app:expand} for the explicit results of this expansion.

One may then wonder whether these large entries in the
anomalous-dimension matrix are an artefact of a badly chosen
renormalisation scheme.  In fact, as discussed for instance in
Refs.~\cite{Buras:1989xd, Herrlich:1994kh}, these entries depend on
the definition of evanescent operators. Needless to say, this scheme
dependence cancels up to the order to which the calculation is
performed.  Nevertheless, the residual scheme dependence can, in
principle, be large.  We have tested this by converting our anomalous
dimension to different schemes chosen such that many of the large
entries vanish.  While the residual {\em scale} dependence is indeed
somewhat smaller in these schemes, the central values of the Wilson
coefficients strongly depend on the choice of scheme, taking values in
approximately the same range as indicated by the scale dependence of
the results in our original calculation.

All these observations hint at a slow convergence of the perturbation
series. We therefore adopt the following prescription for the
numerical values of the Wilson coefficients, including our best
estimate of the associated remaining theory uncertainty: we obtain the
``central values'' of the Wilson coefficients as the value for
$\muew=M_h$ and $\mu_{b(c)}=m_{b(c)}(m_{b(c)})$ and assign as the
theory uncertainty either half the range of the NLL scale variations,
or half the shift between LL and NLL, whichever is larger.

\begin{table}
  \centering
  \begin{tabular}{crr}
            & \multicolumn{1}{c}{\bf Bottom Case}               & \multicolumn{1}{c}{\bf Charm Case}  \\\hline\hline
    $a_{3u}\pm \Delta a_{3u}$ & $ 0.125\pm0.034$                & $( 6.78\pm2.00)\times10^{-3}$ \\
    $a_{3d}\pm \Delta a_{3d}$ & $ 0.115\pm0.028$                & $( 1.55\pm0.49)\times10^{-2}$ \\
    $a_{4} \pm \Delta a_{4}$  & $-0.311\pm0.076$                & $(-6.25\pm1.58)\times10^{-2}$ \\
    $a_{w} \pm \Delta a_w$    & $(4.10\pm 2.50)\times 10^{-3}$  & $( 4.10\pm2.05)\times10^{-3}$ \\[0.5em]
    $b_{3} \pm \Delta b_3$    & $(-1.66\pm1.35)\times10^{-4}$   & $(-2.00\pm3.68)\times10^{-5}$ \\
    $b_{4} \pm \Delta b_4$    & $( 1.84\pm0.36)\times10^{-3}$   & $(-2.71\pm3.36)\times10^{-4}$ \\
    $b_{w} \pm \Delta b_w$    & $(-1.63\pm1.00)\times10^{-2}$   & $(-3.41\pm1.69)\times10^{-3}$ \\\hline
  \end{tabular}
  \caption{Numerical values for the coefficients $a$ and $b$ and their
    respective uncertainties as defined in Eq.~\eqref{eq:C:numbers}.
    \label{tab:magic}}
\end{table}

We then find the three-flavour Wilson coefficients, evaluated at
$2$\,GeV, of electric dipole, chromoelectric dipole, and Weinberg
operators to be:
\begin{equation}
  \label{eq:C:numbers}
  \begin{split}
  C_{3,f=3}^u(2\,\text{GeV}) &= \left[
    (a_{3u} \pm \Delta a_{3u}) \kappa_q   \sin\phi_q
  + (b_{3} \pm \Delta b_{3}) \kappa_q^2 \sin\phi_q \cos\phi_q \right] \frac{\text{GeV}^2}{M_h^2}\,,\\
  C_{3,f=3}^{d,s}(2\,\text{GeV}) &=  \left[
    (a_{3d} \pm \Delta a_{3d}) \kappa_q \sin\phi_q
  + (b_{3} \pm \Delta b_{3}) \kappa_q^2 \sin\phi_q \cos\phi_q \right] \frac{\text{GeV}^2}{M_h^2}\,,\\
  C_{4,f=3}^{u,d,s}(2\,\text{GeV}) &= \left[
	  (a_{4} \pm \Delta a_{4})\kappa_q   \sin\phi_q
	+ (b_{4} \pm \Delta b_{4})\kappa_q^2 \sin\phi_q \cos\phi_q\right] \frac{\text{GeV}^2}{M_h^2}\,,\\
  C_{w,f=3}(2\,\text{GeV}) &= \left[
	 (a_{w} \pm \Delta a_{w})\kappa_q   \sin\phi_q
    +(b_{w} \pm \Delta b_{w})\kappa_q^2 \sin\phi_q \cos\phi_q \right]\frac{\text{GeV}^2}{M_h^2}\,.
  \end{split}
\end{equation}
Here, the subscripts $q=b,c$ refer to the cases of the bottom and
charm quark, respectively. The values for the $a$ and $b$ coefficients
and their respective uncertainties are given, for the two cases, in
the second and third columns of table~\ref{tab:magic}.
We see that in both cases the residual uncertainty on the dipole
Wilson coefficients is of the order of $30\%$, while the contributions
induced by the Weinberg operator have larger uncertainties. They could
potentially be reduced by a next-to-next-to-leading-logarithmic
calculation.

\boldmath
\subsection{Bounds on CP phases from hadronic EDMs}
\unboldmath

\begin{figure}[ht!]
	\centering
	\includegraphics{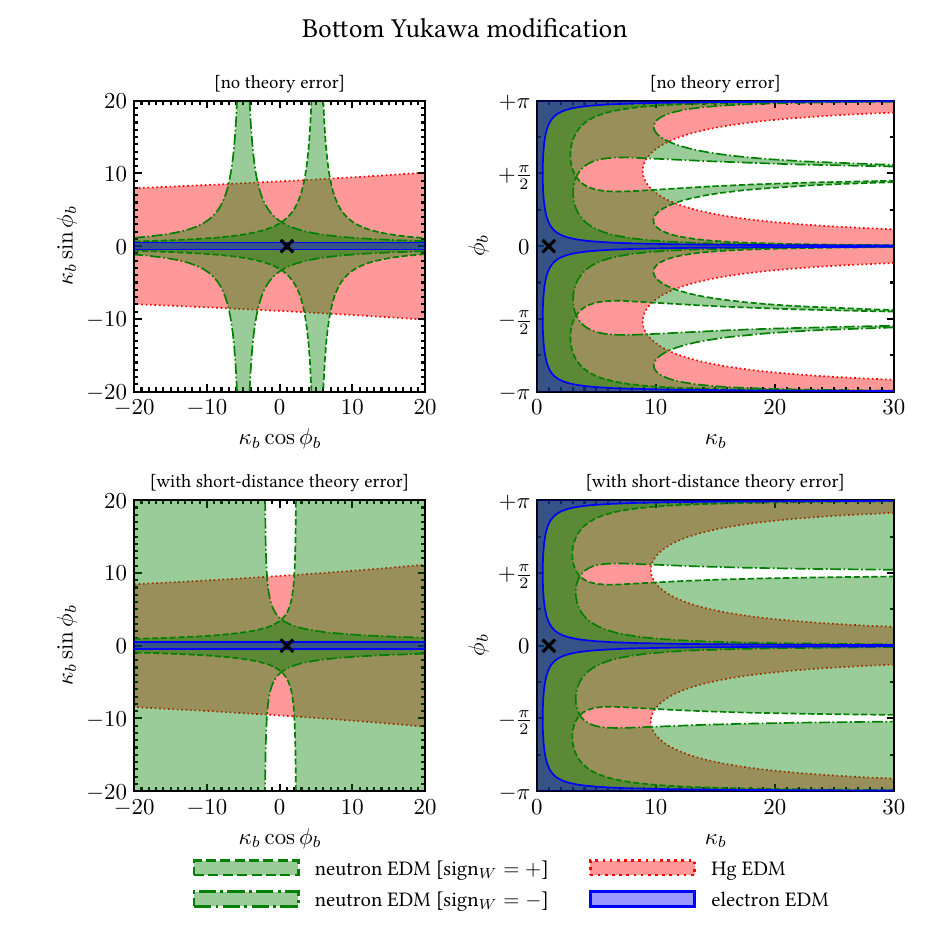}
	\caption{EDM constraints on anomalous CP violating
          bottom-quark Yukawas.  Coloured are the allowed 68.26\% CL
          regions for the two-parameter space (left:
          $\kappa_b\sin\phi_b$ and $\kappa_b\cos\phi_b$, right:
          $\kappa_b$ and $\phi_b$).  In green, the allowed region from
          the neutron EDM for the case of negative (dashed-dotted
          line) and positive (dashed line) sign for the
          Weinberg-operator contribution.  In red (dotted line) the
          allowed region from the mercury EDM, and in blue (solid
          line) the one from the electron EDM.
	\label{fig:EDMbottom}}
\end{figure}

In this section we derive constraints on new CP-violating phases in
the bottom or charm Yukawa under the simplifying assumption that only
one such phase is present. A global fit for the general case including
constraints from the LHC will be presented in a future
publication~\cite{BCSS2019}.

We start with a generic parameterisation of a nuclear dipole moment, $X$, as
\begin{equation}
	d_X = d_X(f_{X,i}, a_i, b_i)\,.
\end{equation}
Here, the $f_{X,i}$ with $i=1,\dots,n$ are the hadronic input
parameters entering the prediction of the given $d_X$. We denote their
uncertainties by $\Delta f_{X,i}$. The $a_i$, with $i=3u,3d,4,w$, and
$b_i$, with $i=3,4,w$, are defined in Section~\ref{sec:wilson} and
given in table~\ref{tab:magic}.  They parameterise the Wilson
coefficients that contribute to $d_X$ if either the bottom or the
charm quark Yukawa is modified.  By standard quadratic error
propagation we compute the total theory uncertainty
\begin{equation}
\Delta d_{X}^{\text{th}} = \biggl(
  \underbrace{\sum_{i=1}^n \left( \frac{\partial d_X}{\partial f_{X,i}} \Delta f_{X,i} \right)^2
             }_{\equiv (\Delta d_X^{\text{hadronic}})^2}
  +
  \underbrace{ \sum_{j=1} \left( \frac{\partial d_X}{\partial a_{j}} \Delta a_{j} \right)^2
              + \sum_{j=1} \left( \frac{\partial d_X}{\partial b_{j}} \Delta b_{j} \right)^2
             }_{\equiv (\Delta d_X^{\text{short-distance}})^2}
\biggr)^{\frac{1}{2}} \,.
\end{equation}
To derive the allowed confidence level (CL) intervals from the
measurements of dipole moments and to combine them  we construct a
$\chi^2$ function of two parameters, $\kappa_q,\phi_q$ or equivalently
of $\kappa_q\sin\phi_q,\,\kappa_q\cos\phi_q$:
\begin{equation}
	\chi^2(\kappa_q,\phi_q) = \sum_X \frac{(d_X^{\text{obs}} - d_X)^2}{(\Delta d_X^{\text{exp}})^2 + (\Delta d_X^{\text{hadronic}})^2 + (\Delta d_X^{\text{short-distance}})^2}\,,
\end{equation}
where we have neglected the tiny SM contribution to any EDM.  The
allowed $68.27\%$ CL region for the two-parameters are then given by
the region $\chi^2(\kappa_q,\phi_q) - \chi^2_{\text{min}} \leq2.30$.

The relation between the coefficients $d_q$, $\tilde d_q$, and $w$ in
Eq.~\eqref{eq:LeffN} and the Wilson coefficients in the three-flavour
EFT is
\begin{equation} \label{eq:ddw}
\begin{split}
d_q (\mu) & = \sqrt{2} G_F \, \frac{Q_q e}{4 \pi \alpha_s} \,
\, m_q \, C_3^q (\mu)
 -12 e Q_q Q_b^2 \, \frac{\alpha}{(4\pi)^3} \sqrt{2}
G_F m_q \, \kappa_b \sin\phi_b \, x_b \left (
\log^2 x_b + \frac{\pi^2}{3} \right ) \,,\\[0.5em]
\tilde d_q (\mu) & = - \sqrt{2} G_F \, \frac{1}{4 \pi \alpha_s}
\, m_q \, C_4^q (\mu) \,,\\
w (\mu) & = \sqrt{2} G_F \, \frac{1}{4 \pi \alpha_s}
\, C_w (\mu) \,.
\end{split}
\end{equation}
In the expression for $d_q$ above we have included the electroweak
contribution of Eq.~\eqref{eq:dsw}. From this we obtain the neutron
EDM as
\begin{equation}
  \frac{d_n}{e} = (1.1 \pm 0.55) (\tilde d_d + 0.5 \, \tilde d_u) + \Big(
  g_T^u \frac{d_u}{e} + g_T^d \frac{d_d}{e} + g_T^s \frac{d_s}{e}
  \Big) \pm (22 \pm 11) \, w \, \text{MeV} \,,
  \label{eq:dn}
\end{equation}
where the matrix elements of the electric dipole operator are
parameterised by $g_T^u = -0.24(3)$, $g_T^d = 0.85(8)$, $g_T^s =
-0.012(18)$. These values are calculated using lattice QCD with $2+1$
active flavors and are converted to the $\overline{\text{MS}}$ scheme
at $2$\,GeV~\cite{Yamanaka:2018uud, Aoki:2019cca} (see
Ref.~\cite{Gupta:2018lvp} for a $N_f=2+1+1$ result).  By using the
lattice values of the three-flavour theory we include the effect of
threshold corrections associated to the charm quark.  To capture this
effect in the four-flavour theory the matrix elements of four-quark
operators with charm quarks would also need to be evaluated via
lattice methods.  The hadronic matrix elements of the chromoelectric
dipole and the Weinberg operators are estimated using QCD sum rules
and chiral techniques~\cite{Pospelov:2005pr,Engel:2013lsa}.  Notice
that the sign of the hadronic matrix element of the Weinberg is not
known, and thus the allowed CL intervals will depend on it.  For
prospects on obtaining the latter via lattice calculations see
Refs.~\cite{Bhattacharya:2015rsa, Bhattacharya:2016rrc}.

The experimental 90\% CL exclusion bound obtained in
Ref.~\cite{Baker:2006ts} is $|d_n| < 2.9 \times 10^{-26} \, e \,
\text{cm}$.  Using the central values of the Wilson coefficients in
table~\ref{tab:magic} and the two-parameter $\chi^2$ we compute the
allowed 68.27\% CL intervals for the bottom- and charm-quark
cases. The label ``sign$_w$'' indicates whether the sign of the
Weinberg-operator contribution in Eq.~\eqref{eq:dn} is taken to be
positive or negative.  We show the CL intervals for the cases in
which: i) no theory uncertainty is included ({\it no theory error}
label), ii) only the short-distance theory uncertainty is included
({\it with short-distance theory error} label) iii) the short-distance
theory uncertainty is added in quadrature with the present theory
uncertainties of the hadronic input ({\it with theory error} label).
For brevity we introduce the short-hand notation
$\sin\phi_{b(c)}\equiv s_{b(c)}$ and $\cos\phi_{b(c)}\equiv c_{b(c)}$.
For the bottom case we find the allowed 68.27\% CL regions to be:
\begin{align}
\kappa_b|s_b|\sqrt{1+0.40 \kappa_b c_b+0.040  \kappa^2_b c^2_b} &\leq 3.5
  & &\text{[sign$_w= -$, no   theory error]}\,,\\
\kappa_b|s_b|\sqrt{1+0.49 \kappa_b c_b+0.0037 \kappa^2_b c^2_b} &\leq 3.8
  & &\text{[sign$_w= -$, with short-distance theory error]}\,,\\
\kappa_b|s_b|\sqrt{1+5.4  \kappa_b c_b-0.27   \kappa^2_b c^2_b} &\leq 12
  & &\text{[sign$_w= -$, with theory error]}\,,\\[0.5em]
\kappa_b|s_b|\sqrt{1-0.39 \kappa_b c_b+0.037  \kappa^2_b c^2_b} &\leq 3.1
  & &\text{[sign$_w= +$, no   theory error]}\,,\\
\kappa_b|s_b|\sqrt{1-0.45 \kappa_b c_b+0.0082 \kappa^2_b c^2_b} &\leq 3.4
  & &\text{[sign$_w= +$, with short-distance theory error]}\,,\\
\kappa_b|s_b|\sqrt{1-1.5  \kappa_b c_b-0.054  \kappa^2_b c^2_b} &\leq 6.4
  & &\text{[sign$_w= +$, with theory error]\,.}
\end{align}
For the charm case we find:
\begin{align}
\kappa_c|s_c|\sqrt{1+0.53 \kappa_c c_c +0.069  \kappa^2_c c^2_c} &\leq 21
 &  &\text{[sign$_w= -$, no   theory error]}\,,\\
\kappa_c|s_c|\sqrt{1+0.78 \kappa_c c_c +0.047  \kappa^2_c c^2_c} &\leq 25
 &  &\text{[sign$_w= -$, with short-distance theory error]}\,,\\[0.5em]
\kappa_c|s_c|\sqrt{1-0.31 \kappa_c c_c +0.024  \kappa^2_c c^2_c} &\leq 13
 &  &\text{[sign$_w= +$, no   theory error]}\,,\\
\kappa_c|s_c|\sqrt{1-0.35 \kappa_c c_c +0.011  \kappa^2_c c^2_c} &\leq 14
 &  &\text{[sign$_w= +$, with short-distance theory error]}\,,\\
\kappa_c|s_c|\sqrt{1-0.61 \kappa_c c_c -0.010  \kappa^2_c c^2_c} &\leq 19
 &  &\text{[sign$_w= +$, with theory error]}\,.
\end{align}
Due to the large theory uncertainties there is no 68.27\% CL interval
for the case sign$_w= -$ when the full theory uncertainties are
included.

Other hadronic EDMs give, in principle, complementary bounds. For
instance, the contribution to the mercury EDM is given
by~\cite{Pospelov:2005pr}
\begin{equation}
  \frac{d_\text{Hg}}{e} = - 1.8 \times 10^{-4} \big(4_{-2}^{+8}\big)
  \big( \tilde d_u - \tilde d_d \big)\,.
	\label{eq:Hgprediction}
\end{equation}
Using the current upper experimental 95\% CL bound $|d_\text{Hg}| <
3.1 \times 10^{-29} \, e \, \text{cm}$ ~\cite{Griffith:2009zz} we
compute the allowed 68.27\% CL intervals from the two-parameter
$\chi^2$.  The presently large hadronic uncertainty in
Eq.~\eqref{eq:Hgprediction} does not constrain the parameter space at
the 68.27\% CL. We thus include only the theory uncertainties
associated to short-distance dynamics in our bounds.
For the bottom case we find the allowed 68.27\% CL regions to be:
\begin{align}
\kappa_b|s_b|\sqrt{1-0.012  \kappa_b c_b+0.000035 \kappa^2_b c^2_b}&\leq8.9
 & &\text{[no   theory error]}\,,\\
\kappa_b|s_b|\sqrt{1-0.014  \kappa_b c_b+0.000037 \kappa^2_b c^2_b}&\leq9.6
 & &\text{[with short-distance theory error]\,.}
\end{align}
For the charm case we find
\begin{align}
\kappa_c|s_c|\sqrt{1+0.0087 \kappa_c c_c+0.000019 \kappa^2_c c^2_c}&\leq44
 & &\text{[no   theory error]}\,,\\
\kappa_c|s_c|\sqrt{1+0.010 \kappa_c c_c-0.000056  \kappa^2_c c^2_c}&\leq48
 & &\text{[with short-distance theory error]\,.}
\end{align}
We see that even if we neglect the present theory uncertainties the
constraints from mercury EDM cannot compete with the ones from the
neutron EDM.

It is instructive to compare the constraints obtained from hadronic
EDMs to the constraints from the bound on the electron EDM, obtained
recently by the ACME collaboration.  The contribution of a modified
bottom Yukawa to the electron EDM can be easily obtained by the
substitutions $Q_q \to Q_e$ and $m_q \to m_e$ in Eq.~\eqref{eq:dsw},
and similarly for a modified charm Yukawa.  Using the ACME result,
$|d_e| < 1.1 \times 10^{-29} \, e \, \text{cm} \quad \text{(90\% CL)}$
~\cite{ACME2018}, we compute the corresponding allowed intervals.  The
electron EDM depends solely on the combination $\kappa_q\sin\phi_q$,
which is thus constrained by a one-parameter $\chi^2$ function to be
$\kappa_b|s_b| \leq 0.32$ and $\kappa_c|s_c| \leq 0.82$ at the
68.26\%\,CL for the bottom- and charm-quark case, respectively.  To
compare with the neutron and mercury EDM allowed two-parameter
68.26\%\,CL intervals we also list the corresponding ones for the
electron EDM:
\begin{align}
	\kappa_b |s_b| & \leq 0.48 &&\text{[no theory error]}\,,\\
	\kappa_c |s_c| & \leq 1.2           &&\text{[no theory error]}\,.
\end{align}
Since for the electron EDM there is no hadronic input, the theory
uncertainties originate solely from higher electromagnetic corrections
and are small.\footnote{Our bound on the bottom Yukawa seems consistent
  with the one recently obtained in
  Ref.~\cite{Egana-Ugrinovic:2018fpy}.} We see that currently the
bound from the electron EDM is stronger than the one from the neutron
EDM.  However, both experimental progress and the anticipated lattice
calculations will strengthen the bounds from neutron and other
hadronic EDMs. The combination of leptonic and hadronic EDMs can also
be used as a strategy to disentangle effect of having multiple
CP-violating Yukawas.

\begin{figure}[ht!]
	\centering
	\includegraphics{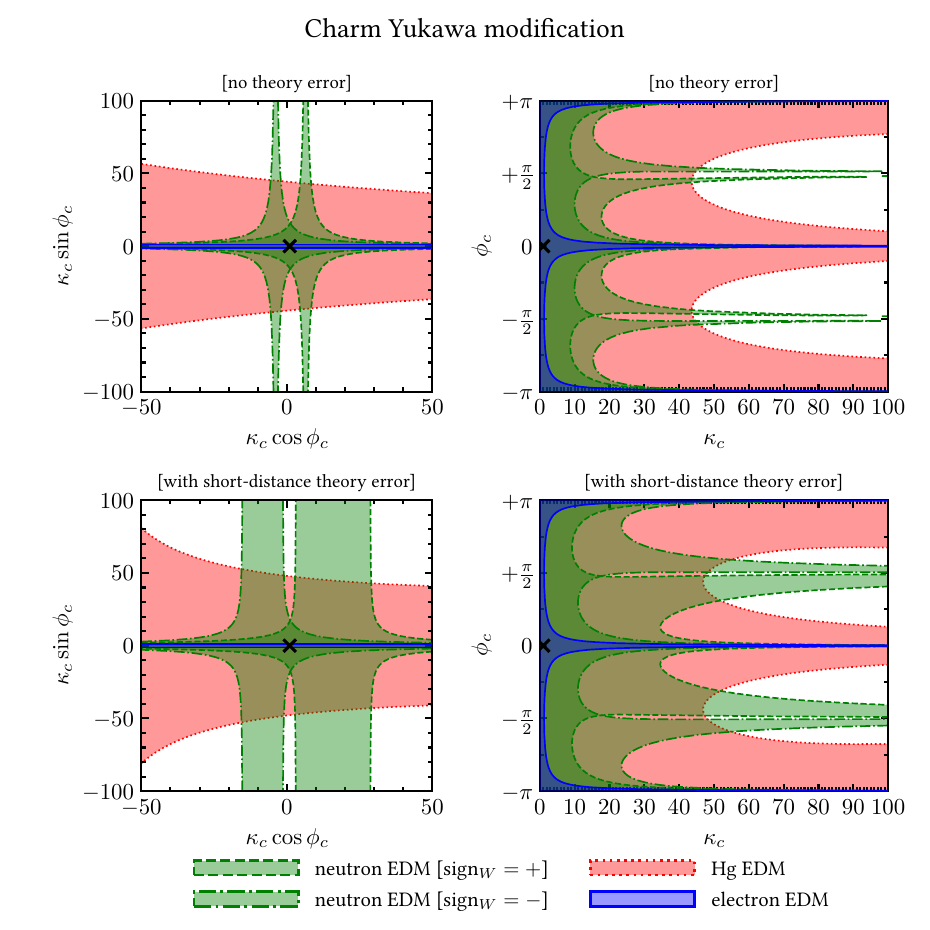}
	\caption{The same as Fig.~\ref{fig:EDMbottom}, but for the case of
		anomalous CP violating charm-quark Yukawas.
	\label{fig:EDMcharm}}
\end{figure}

We illustrate the results of this section for the bottom- and
charm-quark cases in Figs.~\ref{fig:EDMbottom} and \ref{fig:EDMcharm},
respectively.  We show in colour the allowed $68.26\%$ CL regions of
the two-parameter space for different EDMs.  In the plots on the left
we take the two parameters to be $\kappa_{b(c)}\sin\phi_{b(c)}$ and
$\kappa_{b(c)}\cos\phi_{b(c)}$; in the plots on the right we choose
the parameters to be $\kappa_{b(c)}$ and $\phi_{b(c)}$.  In the upper
plots we have included no theory uncertainties; in the lower ones we
folded in the theory uncertainties associated to short-distance
dynamics.

\section{Conclusions}
\label{sec:conclusions}

We presented the complete two-loop QCD anomalous-dimension matrix for
the mixing of CP-odd scalar and tensor operators in an EFT valid below
the electroweak-symmetry breaking scale. We used the results to
perform a next-to-leading-logarithmic RG analysis of the Wilson
coefficients from the weak scale to the hadronic scale of $2$\,GeV,
calculating also the requisite finite matching corrections at the
heavy-flavor thresholds.

We applied our calculation to a new-physics scenario where new,
flavor-conserving, CP-violating phases appear in the Higgs Yukawa
couplings to the bottom or charm quark.  We calculated the initial
conditions at the weak scale up to NLO, and solved the RG equations to
compute the induced coefficients of the CP-violating electric and
chromoelectric dipole operators and the Weinberg operator in the
three-flavor EFT at the hadronic scale $2$\,GeV.

We find large shifts, as well as a large residual scale and scheme
dependence of the dipole Wilson coefficients at
next-to-leading-logarithmic order. We interpret this as a hint of a
slowly converging perturbation series.

The dipole and Weinberg operators contribute to the electric dipole
moment of the neutron and mercury. Assuming a Peccei--Quinn-type
solution to the strong CP problem we can then derive constraints on
the modified Yukawa couplings from the experimental bound on the
neutron EDM.  These constraints are currently not as strong as those
derived from measurements of the electron EDM, but will play an
important role in future global fits to modified Higgs Yukawas. This
is true in particular in light of the progress expected in lattice
calculations of the hadronic matrix elements, and future improvements
regarding the experimental bounds on various hadronic EDMs.

\section*{Acknowledgments}
We thank Marco Ciuchini, Martin Gorbahn, Mikolai Misiak, and Michael
Ramsey-Musolf for useful discussions, and the JHEP referee for
pointing out the inconsistency of our treatment of $\gamma_5$ in the
previous versions of this paper. This work was performed in part at
Aspen Center for Physics, which is supported by National Science
Foundation grant PHY-1607611.  E.~S. is supported by the Fermi
Fellowship at the Enrico Fermi Institute and by the U.S.\ Department
of Energy, Office of Science, Office of Theoretical Research in High
Energy Physics under Award No.\ DE-SC0009924
and by the Swiss National Science Foundation under contract 200021--178999.
J.~B. acknowledges
support in part by DoE grants DE-SC0020047 and DE-SC0011784.

\appendix

\section{Unphysical operators}
\label{app:unphys}

Unphysical operators enter our calculation in two different ways. They
are needed in order to project all off-shell Greens functions, and
they arise as counterterms in intermediate steps of the
calculation. They are called ``unphysical'' because they vanish either
via the equations of motion (e.o.m.)  of the quark fields for onshell
external states, or via algebraic relations that are valid in $d=4$,
but not in $d\neq 4$.

\subsection{E.o.m.-vanishing operators}
These operators have matrix-elements that vanish via the e.o.m.\ of the quark
field. The following two gauge-invariant ones enter our computation at the two-loop
level:
\begin{equation} \label{eq:eom:LO}
\begin{split}
N_1^q
 & = \frac{m_q}{2g_s^2} \bar q
     \bigl[ \stackrel{\leftarrow}{\slashed{D}}\stackrel{\leftarrow}{\slashed{D}} i\gamma_5
            + i\gamma_5 \slashed{D}\slashed{D}\bigr] q \,,\\
N_2^q
 & = \frac{m_q}{2g_s^2} \bar q
     \bigl[   \stackrel{\leftarrow}{\slashed{D}} \stackrel{\leftarrow}{D^\sigma}
              \gamma^\mu\gamma^\nu\gamma^\rho
            - \gamma^\mu\gamma^\nu\gamma^\rho D^\sigma \slashed{D} \bigr]q \epsilon_{\mu\nu\rho\sigma}\,,\\
\end{split}
\end{equation}
Additionally, the following operators, which are not gauge-invariant, are also required
in intermediate steps to determine all counterterms and project the off-shell amplitudes
\begin{equation} \label{eq:eom:LO:gauge-variant}
\begin{split}
N_{\gamma1}^q
 & = -\frac{im_q e Q_q}{2 g_s^2} \bar q
      \bigg[\stackrel{\leftarrow}{\slashed{D}} \slashed{A} i \gamma_5
            - i \gamma_5\slashed{A} \slashed{D} \bigg]q  \,, \\
N_{\gamma2}^q
 & = -\frac{im_q e Q_q}{2 g_s^2} \bar q
     \bigg[\stackrel{\leftarrow}{\slashed{D}}A^{\sigma}\gamma^\mu\gamma^\nu\gamma^\rho
           +\gamma^\mu\gamma^\nu\gamma^\rho A^{\sigma}\slashed{D} \bigg]
           q \epsilon_{\mu\nu\rho\sigma}\,, \\
N_{g1}^q
 & = +\frac{im_q }{2g_s}
      \bar q \bigg[\stackrel{\leftarrow}{\slashed{D}} \slashed{G}i\gamma_5
      -  i \gamma_5\slashed{G} \slashed{D}\bigg] q\,, \\
N_{g2}^q
 & = +\frac{im_q }{2g_s}
      \bar q \bigg[\stackrel{\leftarrow}{\slashed{D}}G^{\sigma}\gamma^\mu\gamma^\nu\gamma^\rho
                   +\gamma^\mu\gamma^\nu\gamma^\rho G^{\sigma}\slashed{D} \bigg]
           q \epsilon_{\mu\nu\rho\sigma}\,, \\
N_{ggg} & = \frac{1}{g_s} f^{abc} G^{\mu,a} \widetilde{G}_{\mu\nu}^b (D_\rho G^{\rho\nu})^c \,.
\end{split}
\end{equation}
The covariant derivative acting on quarks is defined as
\begin{equation} \label{eq:codev}
D_\mu \equiv \partial_\mu - i g_s T^a G_\mu^a  + i e Q_q A_\mu \,,
\end{equation}
with $Q_q$ the quark electrical charge. Accordingly, the gluon
field-strength tensor is
\begin{equation} \label{eq:gluonG}
G_{\mu\nu}^a \equiv \partial_{\mu} G_{\nu}^a - \partial_{\nu}
G_{\mu}^a + g_s \, f^{abc} \, G_{\mu}^b G_{\nu}^c \,.
\end{equation}
The covariant derivative acting on color octets is given by
\begin{equation} \label{eq:codev:oct}
D_\mu^{ab} \equiv \partial_\mu \delta^{ab} - g_s f^{abc} G_\mu^c \,.
\end{equation}

\subsection{Evanescent operators}
Next we list the evanescent operators that enter our computation at one and two-loop order.
The leading-order anomalous dimension does not depend on their definition, but the
next-to-leading-order one does.

In the $q$--$q$--$A$ sector we require the operator
\begin{equation} \label{eq:evan:qA:larin}
\begin{split}
  {E}_{\gamma }^{q} &= \frac{eQ_q}{4} \frac{m_q}{g_s^2}
            \bar q \{\sigma^{\mu\nu}, i \gamma_5\} q \, F_{\mu\nu} - Q_{3}^q\,.
\end{split}
\end{equation}

Analogously, in the $q$--$q$--$G$ sector we require the operator
\begin{equation} \label{eq:evan:qG:larin}
\begin{split}
  {E}_{g }^{q} &= -\frac{1}{4} \frac{m_q}{g_s}
            \bar q \{\sigma^{\mu\nu}, i \gamma_5\} T^a q \, G^a_{\mu\nu} - Q_{4}^q\,.
\end{split}
\end{equation}

Notice that in Eqs.~\eqref{eq:evan:qA:larin} and~\eqref{eq:evan:qG:larin}, as well as in those that follow,
we have defined the $\gamma$-algebra structure in terms of anticommutators with $\gamma_5$, i.e., $\{\Gamma, i\gamma_5\}$.
This ensures that the operators are self-conjugate not just in $d=4$ dimensions but also
in $d=4-2\epsilon$ dimensions.

The evanescent operators required in the $q$--$q'$ sector read
\begin{equation} \label{eq:evan:LO}
\begin{split}
{E}_1^{qq'} &=\frac{1}{2}
(\bar q   \gamma_{[\mu}\gamma_{\nu]} q) \,
(\bar q'  \{\gamma^{[\mu}\gamma^{\nu]}, i\gamma_5\} q')
                  + O_3^{qq'}\,,\\
{E}_2^{qq'} &=\frac{1}{2}
(\bar q   \gamma_{[\mu}\gamma_{\nu]} T^a q) \,
(\bar q'  \{\gamma^{[\mu}\gamma^{\nu]}, i\gamma_5\} T^a q')
                  + O_4^{qq'}\,,\\
{E}_{3}^{qq'} &=
\frac{}{}
\big[
(\bar q  \gamma_{[\mu}\gamma_{\nu]}\gamma^{[\rho}\gamma^{\sigma]} q) \,
(\bar q' \gamma^{[\mu}\gamma^{\nu]}\gamma^{[\tau}\gamma^{\upsilon]} q')
+
(\bar q  \gamma^{[\rho}\gamma^{\sigma]}\gamma_{[\mu}\gamma_{\nu]} q) \,
(\bar q' \gamma^{[\tau}\gamma^{\upsilon]}\gamma^{[\mu}\gamma^{\nu]} q')
\big] \epsilon_{\rho\sigma\tau\upsilon}\\
&\quad
-48 (O_1^{qq'}+O_1^{q'q})
+16 O_3^{qq'}\,,\\
{E}_{4}^{qq'} &=
\frac{1}{2}
\big[
(\bar q  \gamma_{[\mu}\gamma_{\nu]}\gamma^{[\rho}\gamma^{\sigma]} T^a q) \,
(\bar q' \gamma^{[\mu}\gamma^{\nu]}\gamma^{[\tau}\gamma^{\upsilon]} T^a q')\\
& \qquad +
(\bar q  \gamma^{[\rho}\gamma^{\sigma]}\gamma_{[\mu}\gamma_{\nu]} T^a q) \,
(\bar q' \gamma^{[\tau}\gamma^{\upsilon]}\gamma^{[\mu}\gamma^{\nu]} T^a q')
\big] \epsilon_{\rho\sigma\tau\upsilon} \\
&\quad
-48 (O_2^{qq'}+O_2^{q'q})
+16 O_4^{qq'}\,,\\
{E}_{5}^{qq'} &= \frac{1}{2}
(\bar q   \gamma_{[\mu}\gamma_{\nu}\gamma_{\rho}\gamma_{\sigma]} q) \,
(\bar q'  \{\gamma^{[\mu}\gamma^{\nu}\gamma^{\rho}\gamma^{\sigma]}, i\gamma_5\} q')
                  -24 O_1^{q'q}\,,\\
{E}_{6}^{qq'} &=\frac{1}{2}
(\bar q   \gamma_{[\mu}\gamma_{\nu}\gamma_{\rho}\gamma_{\sigma]} T^a q) \,
(\bar q'  \{\gamma^{[\mu}\gamma^{\nu}\gamma^{\rho}\gamma^{\sigma]}, i\gamma_5\} T^a q')
                  -24 O_2^{q'q}\,,\\
{E}_{7}^{qq'} &=\frac{1}{2}
(\bar q   \gamma_{[\mu}\gamma_{\nu}\gamma_{\rho}\gamma_{\sigma}\gamma_{\tau}\gamma_{\upsilon]} q) \,
(\bar q'  \{\gamma^{[\mu}\gamma^{\nu}\gamma^{\rho}\gamma^{\sigma}\gamma^{\tau}\gamma^{\upsilon]}, i\gamma_5\} q') \,,\\
{E}_{8}^{qq'} &= \frac{1}{2}
(\bar q   \gamma_{[\mu}\gamma_{\nu}\gamma_{\rho}\gamma_{\sigma}\gamma_{\tau}\gamma_{\upsilon]} T^a q) \,
(\bar q'  \gamma^{[\mu}\gamma^{\nu}\gamma^{\rho}\gamma^{\sigma}\gamma^{\tau}\gamma^{\upsilon]}, i\gamma_5\} T^a q') \,,\\
{E}_{9}^{qq'} &=
\frac{1}{2}
\big[
(\bar q  \gamma_{[\mu}\gamma_{\nu}\gamma_{\rho}\gamma_{\sigma]}\gamma^{[\tau }\gamma^{\upsilon]} q) \,
(\bar q' \gamma^{[\mu}\gamma^{\nu}\gamma^{\rho}\gamma^{\sigma]}\gamma^{[\zeta}\gamma^{\xi    ]} q')\\
&\qquad +
(\bar q  \gamma^{[\tau }\gamma^{\upsilon]}\gamma_{[\mu}\gamma_{\nu}\gamma_{\rho}\gamma_{\sigma]} q) \,
(\bar q' \gamma^{[\zeta}\gamma^{\xi    ]}\gamma^{[\mu}\gamma^{\nu}\gamma^{\rho}\gamma^{\sigma]} q')
\big]\epsilon_{\tau\upsilon\zeta\xi}
+ 48 O_3^{qq'} \,,\\
{E}_{10}^{qq'} &=
\frac{1}{2}
\big[
(\bar q  \gamma_{[\mu}\gamma_{\nu}\gamma_{\rho}\gamma_{\sigma]}\gamma^{[\tau }\gamma^{\upsilon]} T^a q) \,
(\bar q' \gamma^{[\mu}\gamma^{\nu}\gamma^{\rho}\gamma^{\sigma]}\gamma^{[\zeta}\gamma^{\xi    ]}  T^a q')\\
&\qquad +
(\bar q  \gamma^{[\tau }\gamma^{\upsilon]}\gamma_{[\mu}\gamma_{\nu}\gamma_{\rho}\gamma_{\sigma]} T^a q) \,
(\bar q' \gamma^{[\zeta}\gamma^{\xi    ]}\gamma^{[\mu}\gamma^{\nu}\gamma^{\rho}\gamma^{\sigma]}  T^a q')
\big] \epsilon_{\tau\upsilon\zeta\xi}
+ 48 O_4^{qq'} \,.
\end{split}
\end{equation}
The square brackets denote antisymmetrisation normalized as
\begin{equation*}
\gamma_{[\mu_1,...,\mu_n]} \equiv \frac{1}{n!}\sum_\sigma (-1)^\sigma \gamma_{\mu_{\sigma(1)}}\ldots \gamma_{\mu_{\sigma(n)}}\,.
\end{equation*}
In the $q$--$q$ sector they read
\begin{equation}
\begin{split}
E^{q}_1 & = (\bar q T^a q) (\bar q i \gamma_5 T^a q)
        + \Big( \frac{1}{4} + \frac{1}{2n_c} \Big) O_1^q + \frac{1}{16} O_2^q\,,
	\\[0.5em]
E^{q}_2 & = \frac{1}{2}(\bar q \sigma^{\mu\nu} T^a q) (\bar q \sigma^{\rho\sigma} T^a q) \epsilon_{\mu\nu\rho\sigma}
  + 3 O_1^q - \Big( \frac{1}{4} - \frac{1}{2n_c} \Big) O_2^q\,,
	\\[0.5em]
{E}_3^{q} &= \frac{1}{2}
(\bar q   \gamma_{[\mu}\gamma_{\nu]} q) \,
(\bar q   \{\gamma^{[\mu}\gamma^{\nu]}, i\gamma_5\} q)
+ O_2^{q}\,,\\[0.5em]
E^{q}_4 & =\frac{1}{2}
           (\bar q \gamma_{[\mu}\gamma_{\nu]} T^a q)
           (\bar q \{\gamma^{[\mu}\gamma^{\nu]}, i\gamma_5\} T^a q)
           -3 O_1^q + \left( \frac{1}{4}-\frac{1}{2 n_c} \right)O_2^q\,,
	\\[0.5em]
{E}_{5}^{q} &=
\frac{1}{2}
\big[
(\bar q \gamma_{[\mu}\gamma_{\nu]}\gamma^{[\rho}\gamma^{\sigma]} q) \,
(\bar q \gamma^{[\mu}\gamma^{\nu]}\gamma^{[\tau}\gamma^{\upsilon]} q)
+
(\bar q \gamma^{[\rho}\gamma^{\sigma]}\gamma_{[\mu}\gamma_{\nu]} q) \,
(\bar q \gamma^{[\tau}\gamma^{\upsilon]}\gamma^{[\mu}\gamma^{\nu]} q)
\big] \epsilon_{\rho\sigma\tau\upsilon}\\
&\quad
   -96 O_1^{q}
   +16 O_2^{q}
   \,,\\[0.5em]
{E}_{6}^{q} &=
\frac{1}{2}
\big[
(\bar q \gamma_{[\mu}\gamma_{\nu]}\gamma^{[\rho}\gamma^{\sigma]} T^a q) \,
(\bar q \gamma^{[\mu}\gamma^{\nu]}\gamma^{[\tau}\gamma^{\upsilon]}T^a q)\\
&\qquad +
(\bar q \gamma^{[\rho}\gamma^{\sigma]}\gamma_{[\mu}\gamma_{\nu]} T^a q) \,
(\bar q \gamma^{[\tau}\gamma^{\upsilon]}\gamma^{[\mu}\gamma^{\nu]} T^a q)
\big]\epsilon_{\rho\sigma\tau\upsilon}\\
&\quad
-24 \left( 1-\frac{2}{n_c} \right)O_1^{q}
+ 2 \left( 5-\frac{4}{n_c} \right) O_2^{q}
   \,,\qquad\\[0.5em]
{E}_{7}^{q} &= \frac{1}{2}
(\bar q   \gamma_{[\mu}\gamma_{\nu}\gamma_{\rho}\gamma_{\sigma]} q) \,
(\bar q   \{\gamma^{[\mu}\gamma^{\nu}\gamma^{\rho}\gamma^{\sigma]}, i\gamma_5\} q)
                  -24 O_1^{q}\,,\\[0.5em]
{E}_{8}^{q} &= \frac{1}{2}
(\bar q   \gamma_{[\mu}\gamma_{\nu}\gamma_{\rho}\gamma_{\sigma]} T^a q) \,
(\bar q   \{\gamma^{[\mu}\gamma^{\nu}\gamma^{\rho}\gamma^{\sigma]}, i\gamma_5\} T^a q)
+6\left( 1+\frac{2}{n_c} \right) O_1^q + \frac{3}{2} O_2^q\,,\\[0.5em]
{E}_{9}^{q} &=\frac{1}{2}
(\bar q   \gamma_{[\mu}\gamma_{\nu}\gamma_{\rho}\gamma_{\sigma}\gamma_{\tau}\gamma_{\upsilon]} q) \,
(\bar q   \{\gamma^{[\mu}\gamma^{\nu}\gamma^{\rho}\gamma^{\sigma}\gamma^{\tau}\gamma^{\upsilon]}, i\gamma_5\} q)
\,,\\[0.5em]
{E}_{10}^{q} &=\frac{1}{2}
(\bar q   \gamma_{[\mu}\gamma_{\nu}\gamma_{\rho}\gamma_{\sigma}\gamma_{\tau}\gamma_{\upsilon]} T^a q) \,
(\bar q   \{\gamma^{[\mu}\gamma^{\nu}\gamma^{\rho}\gamma^{\sigma}\gamma^{\tau}\gamma^{\upsilon]}, i\gamma_5\} T^a q)
\,,\\[0.5em]
{E}_{11}^{q} &=
\frac{1}{2}
\big[
(\bar q \gamma_{[\mu}\gamma_{\nu}\gamma_{\rho}\gamma_{\sigma]}\gamma^{[\tau }\gamma^{\upsilon]} q) \,
(\bar q \gamma^{[\mu}\gamma^{\nu}\gamma^{\rho}\gamma^{\sigma]}\gamma^{[\zeta}\gamma^{\xi     ]} q)\\
&\qquad +
(\bar q \gamma^{[\tau }\gamma^{\upsilon]}\gamma_{[\mu}\gamma_{\nu}\gamma_{\rho}\gamma_{\sigma]} q) \,
(\bar q \gamma^{[\zeta}\gamma^{\xi     ]}\gamma^{[\mu}\gamma^{\nu}\gamma^{\rho}\gamma^{\sigma]} q)
\big] \epsilon_{\tau\upsilon\zeta\xi}\\
&\quad
+ 48 O_2^{q}\\
{E}_{12}^{q} &=
\frac{1}{2}
\big[
(\bar q \gamma_{[\mu}\gamma_{\nu}\gamma_{\rho}\gamma_{\sigma]}\gamma^{[\tau }\gamma^{\upsilon]} T^a q) \,
(\bar q \gamma^{[\mu}\gamma^{\nu}\gamma^{\rho}\gamma^{\sigma]}\gamma^{[\zeta}\gamma^{\xi     ]} T^a q) \\
&\qquad +
(\bar q \gamma^{[\tau }\gamma^{\upsilon]}\gamma_{[\mu}\gamma_{\nu}\gamma_{\rho}\gamma_{\sigma]} T^a q) \,
(\bar q \gamma^{[\zeta}\gamma^{\xi     ]}\gamma^{[\mu}\gamma^{\nu}\gamma^{\rho}\gamma^{\sigma]} T^a q)
\big]\epsilon_{\tau\upsilon\zeta\xi}\\
&\quad
-144 O_1^q+ 12\left( 1-\frac{2}{n_c} \right) O_2^{q}\,.
\end{split}
\end{equation}
In simplifying the color algebra, we use the following standard
relation for the generators of SU$(n_c)$:
\begin{equation}
  \sum_a T_{ij}^a T_{kl}^a = \frac{1}{2} \delta_{il} \delta_{jk} - \frac{1}{2n_c} \delta_{ij} \delta_{kl}
\end{equation}
Consequently, Fierz relations on the Lorentz structures, valid in
$d=4$, need to be applied on the operators with $T^a$'s to show that
they are evanescent.

\subsection{Operators related to the infrared rearrangement}
The last class of unphysical operators arises because our infrared
rearrangement breaks gauge invariance in intermediate steps of the
calculation. At the renormalizable level this method generates one
gauge-variant operator corresponding to a gluon-mass term, i.e.,
\begin{equation}
	\Lag \supset \frac{1}{2} Z_{\text{IRA}} m_{\text{IRA}}^2 G_\mu^a G^{\mu,\,a} \,.
\end{equation}
The ``gluon mass'', $m_{\text{IRA}}$, is completely artificial and
drops out of all physical results, and $Z_{\text{IRA}}$ is an
additional renormalisation constant~\cite{Chetyrkin:1997fm}. At the
non-renormalizable level the one-loop insertions of the dimension-five
and dimension-six operators can also induce gauge-invariant operators
that are relics of the infrared rearrangement. For our calculation,
the only relevant ones are the following three
\begin{equation}
  \begin{split}
    P^q          &= m_q \frac{m_{\text{IRA}}^2}{g_s^2} \bar q i \gamma_5 q \,,\\
    P^q_{\gamma} &= \frac{e Q_q m_{\text{IRA}}^2}{2g_s^2} \bar q \{\slashed{A},i\gamma_5\} q \,,\\
    P^q_{g}      &=     -\frac{m_{\text{IRA}}^2}{2g_s} \bar q \{\slashed{G},i\gamma_5\} q \,.
  \end{split}
\end{equation}

\section{Renormalisation constants}\label{app:Z}
To obtain the two-loop anomalous dimension of the physical sector we
need certain one-loop renormalisation constants involving unphysical
operators. We collect them in this appendix.  We use the following
standard notation for their expansion in $\alpha_s$ and $\epsilon$
\begin{equation}
  Z_{x\to y} = \sum_k \sum_{l=1}^k \frac{\alpha_s^{k}}{(4\pi)^{k} \epsilon^l}
  Z_{x\to y}^{(k,l)} \,.
\end{equation}
In $\overline{\text{MS}}$, the mixing of evanescent operators into
physical also includes finite terms \cite{Dugan:1990df}, thus in this
case the $\epsilon$ expansion starts with $l=0$. The subscripts $x$
and $y$ symbolize sets of Wilson coefficients, for which we use the
following notation and standard ordering:
\begin{equation}
	\begin{split}
q = \{
&C_1^{q},\,C_2^{q},\,C_3^{q},\,C_4^{q}\}\,, \\
qq' = \{
&C_1^{qq'},\, C_2^{qq'},\, C_1^{q'q},\, C_2^{q'q},\, C_3^{qq'},\, C_4^{qq'}\}\,, \\
E^q = \{
& C_{E_1^{q}},\,C_{E_2^{q}},\,C_{E_4^{q}},\,C_{E_{6}^{q}}\}\,,\\
E^{qq'} = \{
  & C_{E_1^{qq'}},\, C_{E_2^{qq'}},\,
    C_{E_1^{q'q}},\, C_{E_2^{q'q}},\,
    C_{E_{3}^{qq'}},\, C_{E_{4}^{qq'}}\}\,, \\
{{E}^q_{\gamma g}} = \{
  & C_{E_{\gamma }^{q}},\,
    C_{E_{g }^{q}}\}\,,\\
P^q = \{ &C_{P^{q}},\,C_{P_{\gamma}^q},\,C_{P_{g}^q}\}\,,\\
N^q = \{  &C_{N_1^{\q}},\,C_{N_2^{\q}},\,C_{N^{\q}_{\gamma 1}},\,C_{N^{\q}_{\gamma 2}},\,C_{N^{\q}_{g1}}\,C_{N^{\q}_{g2}}\}\,.
	\end{split}
\end{equation}

The first necessary input is the mixing of the physical operators into
all the evanescent operators that are generated at one-loop. Using the
same subscript notation as above, the renormalisation constants read
  \begin{equation}
    \begin{split}
Z^{(1,1)}_{\q\to E^\q} &=
\begin{pmatrix}
 0& 0&1& 0\\
 0& -8& 0& - \frac{1}{2}\\
 0 & 0 & 0 & 0 \\
 0 & 0 & 0 & 0
\end{pmatrix}\,,
\hspace{3em}
Z^{(1,1)}_{\q\qp\to E^{\q\qp}} =
\begin{pmatrix}
 0&1& 0& 0& 0& 0\\
\frac{2}{9}&\frac{5}{12}& 0& 0& 0& 0\\
 0& 0& 0&1& 0& 0\\
 0& 0&\frac{2}{9}&\frac{5}{12}& 0& 0\\
 0& 0& 0& 0& 0& - \frac{1}{2}\\
 0& 0& 0& 0& - \frac{1}{9}& - \frac{5}{24}
\end{pmatrix}\,,\\
Z^{(1,1)}_{\q\to {E}_{\gamma g}^\q} &=
\begin{pmatrix}
 2              &2              \\
 24             &24             \\
 0              &0              \\
 -\frac{16}{3}  &\frac{11}{3}   \\
\end{pmatrix}\,.
  \end{split}
  \end{equation}
The remaining mixings of physical operators into evanescent operators
are all zero at one-loop. Furthermore, the finite part of the mixing
of evanescent into physical operators is subtracted by finite
counterterms~\cite{Dugan:1990df}. They read
\begin{equation}
  \begin{split}
Z^{(1,0)}_{E^\q\to\q} &=
\begin{pmatrix}
 -\frac{5}{8} & -\frac{3}{32}      & 0                &0 \\
 -\frac{9}{2} & \frac{1}{8}        & 0                &0 \\
 \frac{169}{36} & -\frac{11}{48}   & -\frac{16}{3}    &\frac{2}{3} \\
 22 & -\frac{235}{18}              & -\frac{128}{3}   &\frac{16}{3} \\
\end{pmatrix}\,,
\hspace{3em}
Z^{(1,0)}_{E^{\q\qp}\to\q\qp} =
\begin{pmatrix}
 0&4& 0& 0& 0& 0\\
\frac{8}{9}&\frac{5}{3}& 0& 0& 0& 0\\
 0& 0& 0&4& 0& 0\\
 0& 0&\frac{8}{9}&\frac{5}{3}& 0& 0\\
 0& 0& 0& 0& 0& - 8\\
 0& 0& 0& 0& - \frac{16}{9}& - \frac{10}{3}
\end{pmatrix}\,,\\
Z^{(1,0)}_{{E}_{\gamma g}^\q\to\q} &=
\begin{pmatrix}
0&0&  0             & 0\\
0&0&  -\frac{8}{9}  & \frac{11}{18}\\
\end{pmatrix}\,.
  \end{split}
\end{equation}
The remaining finite mixings of evanescent into physical operators are
zero.

Furthermore, we need the mixing constants of the physical operators
into the operators arising from infrared rearrangement; they are found
to be
\begin{align}
	&Z^{(1,1)}_{\q\to P^\q}
	=
\begin{pmatrix}
-10 & -2 & -2 \\
24 & -24 & -24 \\
0 & 0 & 0 \\
-16 & 0 & 0
\end{pmatrix}
\,,
&Z^{(1,1)}_{\q\qp\to P^\q } &=
\begin{pmatrix}
 0 & 0 & 0 \\
 0 & 0 & 0 \\
 -12\frac{m_\qp}{m_\q} & 0 & 0 \\
 0 & 0 & 0 \\
 0 & 0 & 0 \\
 0 & 0 & 0 \\
\end{pmatrix}\,.&
\end{align}
All other mixing constants of physical into the IRA
operators are zero.
Special care must be taken to obtain the mixings like $Z^{(1,1)}_{\q\qp\to P^\qp}$.
Apart from the obvious $q\leftrightarrow\qp$ interchange also the ordering
of the operators in the collective blocks is relevant. For example
\begin{equation*}
Z^{(1,1)}_{\q\qp\to P^\qp } =
\begin{pmatrix}
 -12\frac{m_\q}{m_\qp} & 0 & 0 \\
 0 & 0 & 0 \\
 0 & 0 & 0 \\
 0 & 0 & 0 \\
 0 & 0 & 0 \\
 0 & 0 & 0 \\
\end{pmatrix}\,.
\end{equation*}

Finally, we need the mixing constants of the physical operators into
the e.o.m.-vanishing operators. They are uniquely fixed by the $\q\to\q$,
$\q\to \q\gamma$, and $\q\to \q g$ Greens functions and we have verified
that their values are consistent with the renormalisation of the
$\q g\to \q\gamma$ and $\q g\to \q g$ Greens functions.
We find
\begin{equation}
Z^{(1,1)}_{\q\to N^\q } =
\begin{pmatrix}
0&0&0&0&0&0\\
0&0&0&0&0&0\\
0&0&0&0&0&0\\
 -\frac{8}{3}& -\frac{2}{9}& 0& 0&  0 & \frac{3}{8}
\end{pmatrix}\,.
\end{equation}

The two-loop anomalous dimension matrix is given in terms of the one-
and two-loop renormalisation constants by
\begin{equation}
\gamma^{(1)} = 4 Z^{(2,1)} - 2 Z^{(1,1)} Z^{(1,0)} - 2 Z^{(1,0)}
Z^{(1,1)} + 2 \beta_0 Z^{(1,0)} \,.
\end{equation}
The quadratic poles of the two-loop diagrams are fixed by the poles of
the one-loop diagrams via
\begin{equation}\label{eq:2loop:rel}
Z^{(2,2)} = \frac{1}{2} Z^{(1,1)} Z^{(1,1)} - \frac{1}{2} \beta_0
Z^{(1,1)} \,,
\end{equation}
where $\beta_0=\frac{11}{3}n_c-\frac{2}{3}\Nf$. As a check of our
calculation, we computed these poles directly and verified that they
satisfy Eq.~\eqref{eq:2loop:rel}.\\

In our calculation we needed various field and mass renormalisation
constants up to two-loop level, and we have calculated them
explicitly. Writing the expansion
\begin{equation}
  Z_r = \sum_k \sum_{l=1}^k \frac{\alpha_s^{k}}{(4\pi)^{k} \epsilon^l}
  Z_r^{(k,l)} \,,
\end{equation}
with $r = q,m,g_s,G,u,m_{\text{IRA}}$ denoting the quark field, quark
mass, strong coupling, gluon field, ghost field, and artificial gluon
mass renormalisation, respectively, we find
\begin{align}
  Z_q^{(1,1)}           & = - \frac{(n_c^2-1)}{2n_c} \xi_g\,,\\
  Z_m^{(1,1)}           & = - \frac{3(n_c^2-1)}{2n_c}\,,\\
  Z_{g_s}^{(1,1)}        & = - \frac{11}{6}n_c + \frac{1}{3} \Nf\,, \\
  Z_{{\text{IRA}}}^{(1,1)} & = - \frac{n_c}{4} (1 + 3\xi_g) - 2 \Nf\,, \\
  Z_G^{(1,1)}           & = \bigg(\frac{13}{6} - \frac{1}{2} \xi_g\bigg) n_c - \frac{2}{3} \Nf\,, \\
	Z_u^{(1,1)}           & = \frac{n_c(3 - \xi_g)}{4}\,,
\intertext{}
  Z_q^{(2,2)}           & = \frac{\left(n_c^2-1\right) \big[n_c^2 (2 \xi_g+3)-\xi_g\big]}{8 n_c^2} \xi_g\,, \\
  Z_m^{(2,2)}           & = -\frac{\left(n_c^2-1\right) \big(9 - 31n_c^2 + 4n_c \Nf\big)}{8 n_c^2}\,, \\
	Z_{g_s}^{(2,2)}           & = \frac{121}{24}n_c^2 - \frac{11}{6}n_c \Nf + \frac{1}{6}\Nf^2\,, \\
  Z_G^{(2,2)}           & = \bigg( \!\!-\frac{13}{8} -\frac{17}{24} \xi_g +\frac{1}{4} \xi_g^2 \bigg) n_c^2
			  + \bigg( \frac{1}{2} +\frac{1}{3} \xi_g \bigg) n_c \Nf\,,
\intertext{}
Z_q^{(2,1)}           & = -\frac{\left(n_c^2-1\right) \big[n_c^2 \left(\xi_g^2+8 \xi_g+22\right)-4 n_c \Nf+3\big]}{16 n_c^2}\,,\\
  Z_m^{(2,1)}           & = \frac{\left(n_c^2-1\right) \big(9 - 203n_c^2 + 20n_c \Nf\big)}{48 n_c^2}\,, \\
	Z_{g_s}^{(2,1)}           & = -\frac{17}{6}n_c^2 - \frac{1}{4n_c} \Nf + \frac{13}{12}n_c \Nf\,, \\
  Z_G^{(2,1)}           & = \bigg( \frac{59}{16} -\frac{11}{16} \xi_g -\frac{1}{8} \xi_g^2 \bigg) n_c^2
                          - \bigg( \frac{7}{4}n_c -\frac{1}{2n_c} \bigg) \Nf\,,
\end{align}
with $\xi_g$ the gauge fixing parameter in generalized $R_\xi$ gauge.
Our renormalisation constants agree with the results in the
literature~\cite{Gambino:2003zm} if one bears in mind that the
original papers contain some typographical errors.

\section{The one-loop anomalous dimensions}\label{app:ADM}

Using the same notation as in the previous section, we decompose the
anomalous dimension matrix of the physical sector in
subblocks. The anomalous dimension matrices $\gamma_{x\to y}$ admit a
perturbative expansion in the strong coupling constant,
\begin{equation}
\gamma = \frac{\alpha_s}{4\pi} \gamma^{(0)} +
\left(\frac{\alpha_s}{4\pi}\right)^2 \gamma^{(1)} + \ldots \,.
\label{eq:gamma:asexp}
\end{equation}
The one-loop anomalous dimension matrix is given in terms of the
renormalisation constants by
\begin{equation}
\gamma^{(0)} = 2 Z^{(1,1)} \,.
\end{equation}
The explicit results read
\begin{align}
	\gamma^{(0)}_{\q\to\q} &=
\begin{pmatrix}
-10 &   -\frac{1}{6} &                               4 &    4                            \\
 40 &   \frac{34}{3} &                           - 112 & - 16                            \\
  0 &              0 & -\frac{34}{3} + \frac{4}{3} \Nf &    0                            \\
  0 &              0 &  \frac{32}{3}                   & -\frac{38}{3} + \frac{4}{3} \Nf \\
\end{pmatrix}\,,\\
\gamma^{(0)}_{\q\qp\to\q} &=
\begin{pmatrix}
0 & 0 & 0 & 0\\
0 & 0 & 0 & 0\\
0 & 0 & 0 & 0\\
0 & 0 & 0 & 0\\
0 & 0 &  - 48 \frac{Q_{q'}}{Q_q} \frac{m_{q'}}{m_q}& 0\\
0 & 0 &  0& - 8 \frac{m_{q'}}{m_q}
\end{pmatrix}\,,\\
	\gamma^{(0)}_{\q\qp\to\q\qp} &=
\begin{pmatrix}
 -16& 0& 0& 0& 0& - 2\\
 0&2& 0& 0& - \frac{4}{9}& - \frac{5}{6}\\
 0& 0& -16& 0& 0& - 2\\
 0& 0& 0&2& - \frac{4}{9}& - \frac{5}{6}\\
 0& - 48& 0& - 48&\frac{16}{3}& 0\\
 - \frac{32}{3}& - 20& - \frac{32}{3}& - 20& 0&-\frac{38}{3}
\end{pmatrix}\,,\\
	\gamma^{(0)}_{W\to\q} &=
\begin{pmatrix}
  0 & 0 & 0 & 6
\end{pmatrix}\,,\\
\gamma^{(0)}_{W\to W} &=  -8+\frac{8}{3}\Nf\,.
\end{align}
All other physical subblocks are zero at one-loop.
Our one-loop results for the physical sector agree with the results in
the literature~\cite{Buras:2000if, Degrassi:2005zd, Hisano:2012cc},
after accounting for the different normalisation of the operators
and the different conventions in the covariant derivative.

\section{Change of renormalisation scheme\label{app:scheme}}

The (three-loop) mixing among the dipole operators $O_3^q$ and $O_4^q$
has already been calculated in the literature~\cite{Gorbahn:2005sa}
(see Ref.~\cite{Misiak:1994zw} for the earlier two-loop result for the
CP-even dipole operators). In this section we show that our results
are consistent with these calculations.

There are three main differences between our calculation and the
earlier ones: (i) the definition of the physical operators differs in
$d$ space-time dimensions, (ii) the calculation in
Ref.~\cite{Gorbahn:2005sa} has effectively been performed in the NDR
scheme whereas ours in the ``Larin'' scheme, and (iii) a different
normalisation of the operators with factors of quark masses and the
strong coupling has been chosen. Here we show that there is a unique
change of the renormalisation scheme that transforms our result into
that of Refs.~\cite{Misiak:1994zw, Gorbahn:2005sa}.

To address (i), we perform a redefinition of our basis of physical
operators as follows:\footnote{The lattice results used in our
  numerics have been converted to the NDR-$\overline{\text{MS}}$
  scheme at the one-loop level~\cite{Yamanaka:2018uud}. If, in the
  future, the two-loop conversion of the lattice results becomes
  available, one should perform the change of scheme discussed in this
  appendix for our analysis.}
  \begin{equation}
    \begin{split}
  {O'}_3^q & \equiv {O}_3^q + E_\gamma^q
             =\frac{eQ_q}{4} \frac{m_q}{g_s^2}
              \bar q \{\sigma^{\mu\nu}, i \gamma_5\} q \, F_{\mu\nu} \,,\\
  {O'}_4^q & \equiv {O}_4^q + E_g^q
             =  -\frac{1}{4} \frac{m_q}{g_s}
                 \bar q \{\sigma^{\mu\nu}, i \gamma_5\} T^a q \, G^a_{\mu\nu} \,.
    \end{split}
  \end{equation}
All other physical operators are unchanged, i.e. $Q_i' \equiv Q_i$.
The resulting operators correspond, {\it in the NDR scheme}, to the
operators used in Refs.~\cite{Misiak:1994zw, Gorbahn:2005sa}, up to a
normalisation (see below).

To address (ii), we mimick the results obtained in the NDR scheme by
redefining the evanescent operator ${E}_g^q$ to be:
\begin{align}
  {E'}_g^q & \equiv   {E}_g^q
                    + \frac{4}{47} \epsilon {O}_3^q
                    - \frac{19}{94} \epsilon {O}_4^q \,,
\end{align}
with all other operators unchanged. The coefficients in front of the
physical operators are uniquely determined by the requirement to
reproduce the anomalous dimensions of the dipole operators in the
literature.

Finally, to address (iii) we take care of the different overall
normalisation of the operators as follows. If we write the shifted
renormalisation constants as $Z^\prime = \rho Z$, where
\begin{equation}
\rho = 1 + \frac{\alpha_s}{4\pi\epsilon} \rho^{(1,1)} + \bigg(
\frac{\alpha_s}{4\pi\epsilon} \bigg)^2 \big( \rho^{(2,2)} + \epsilon
\rho^{(2,1)} \big) \,,
\end{equation}
then the shifted ADMs are given by
\begin{equation}
\gamma^{\prime(0)} = \gamma^{(0)} + 2 \rho^{(1,1)} \,, \qquad
\gamma^{\prime(1)} = \gamma^{(1)} + 4 \rho^{(2,1)} \,.
\end{equation}
Combining this shift with the change of physical and evanescent
operators specified above (see, e.g., Ref.~\cite{Brod:2010mj} for the
general expressions), we can reproduce both the one- and two-loop
mixing of the dipole operators in Ref.~\cite{Misiak:1994zw,
  Gorbahn:2005sa}. The mixing of four-fermion into dipole operators,
on the other hand, cannot be calculated in the NDR scheme.
For completeness, we provide the full physical two-loop anomalous
dimension matrix in the primed basis:
\begin{align}
\gamma^{(1)}_{\q\qp\to\q} &=
\begin{pmatrix}
 0 & 0 & \frac{32}{27}\frac{m_\qp}{m_\q} & -\frac{640}{47}\frac{m_\qp}{m_\q} \\[0.5em]
 0 & 0 & \frac{40}{141}\frac{m_\qp}{m_\q}-16\frac{Q_\qp}{Q_\q}\frac{m_\qp}{m_\q}
       & -\frac{800}{141}\frac{m_\qp}{m_\q} \\[0.5em]
 0 & 0 & \frac{32}{27}\frac{m_\qp}{m_\q} & -\frac{264}{47}\frac{m_\qp}{m_\q} \\[0.5em]
 0 & 0 & \frac{40}{141}\frac{m_\qp}{m_\q}-\frac{16}{3}\frac{Q_\qp}{Q_\q}\frac{m_\qp}{m_\q}
       & -\frac{110}{47}\frac{m_\qp}{m_\q} \\[0.5em]
 0 & 0 & -448\frac{Q_\qp}{Q_\q}\frac{m_\qp}{m_\q} & 0 \\[0.5em]
 0 & 0 & \frac{2976}{47} \frac{m_\qp}{m_\q} & - \frac{20932}{141} \frac{m_\qp}{m_\q}
\end{pmatrix}\,,\\[1em]
\gamma^{(1)}_{\q\to\q} &=
\begin{pmatrix}
 65-6 \Nf & \frac{\Nf}{54}-\frac{19}{12} & \frac{26896}{1269} & \frac{81211}{1269} \\[0.5em]
 60-\frac{40 \Nf}{9} & \frac{403}{3}-\frac{226 \Nf}{27} & -\frac{339776}{423} & -\frac{65780}{423} \\[0.5em]
 0 & 0 & \frac{460 \Nf}{27}+\frac{100}{9} & 0 \\[0.5em]
 0 & 0 & \frac{368}{3}-\frac{224 \Nf}{27} & -\frac{380 \Nf}{9}+\frac{623}{27}
\end{pmatrix}\,.
\end{align}
The matrix $\gamma^{(1)}_{\q\qp\to\q\qp}$ does not change by our
redefinition of operators.

\section{Expanding the renormalisation group}\label{app:expand}

To gain a better understanding of our results, and as an additional
check of our calculation, we expand the full solution of the RG equations
about the bottom-quark threshold (the procedure for the charm quark is
analogous).

Keeping only the leading nonvanishing terms and including the QED
contribution, Eq.~\eqref{eq:dsw}, we find for the electric dipole
\begin{equation}
	\begin{split}
		d_q  =& \sqrt{2} G_F \frac{Q_q e m_q}{g_s^2} C_3^q (m_b) \\
		     \simeq& \sqrt{2} G_F \frac{Q_q e m_q}{g_s^2} \bigg\{
	-12 Q_b^2 \,
  \frac{\alpha\alpha_s(m_b)}{(4\pi)^2} \, \kappa_b \sin\phi_b \, x_b
  \left ( \log^2 x_b + \frac{\pi^2}{3} \right ) \\
		+&\bigg(\frac{\alpha_s(m_b)}{4\pi}\bigg)^3 \bigg(\\
		&+ C_1^{\q\qp(0)} (M_h) \frac{\gamma_{1,\q\qp \to 4,\q\qp}^{(0)}
    \gamma_{4,\q\qp \to 4,\q}^{(0)} \gamma_{4,\q \to 3,\q}^{(0)}}{48}
  \log^3 x_b \\
		&+ C_1^{\q\qp(0)} (M_h)
    \frac{\gamma_{1,\q\qp \to 4,\q\qp}^{(0)} \gamma_{4,\q\qp \to
        3,\q}^{(1)}+ \gamma_{1,\q\qp \to
        4,\q}^{(1)} \gamma_{4,\q \to 3,\q}^{(0)} + \gamma_{1,\q\qp \to
        3,\q\qp}^{(1)} \gamma_{3,\q\qp \to 3,\q}^{(0)}}{8}\log^2 x_b
    \\
		&+ C_4^{\q\qp(1)} (M_h) \frac{\gamma_{4,\q\qp
        \to 4,\q}^{(0)} \gamma_{4,\q \to 3,\q}^{(0)}}{8}\log^2 x_b
		\bigg)\bigg\} \\
		=& \sqrt{2}
		G_F Q_q e m_q \, \kappa_b \sin\phi_b \, x_b \\
		&\quad \times
  \bigg\{ -12 Q_b^2 \, \frac{\alpha}{(4\pi)^3} \left ( \log^2 x_b +
  \frac{\pi^2}{3} \right ) - \frac{32}{9}
  \frac{\alpha_s^2(m_b)}{(4\pi)^4} \log^3 x_b + \frac{32}{3}
  \frac{\alpha_s^2(m_b)}{(4\pi)^4} \log^2 x_b \bigg\} \,,
	\end{split}
\end{equation}
where the superscripts ``$(0)$'' and ``$(1)$'' denote the tree-level
and one-loop contributions to the initial conditions of the Wilson
coefficients at the weak scale (we omit here the small logarithmic
contributions $\propto \log(\muew^2/M_h^2)$). Using $\alpha_s(m_b)
\sim 0.22$, the ratio of QED to LL to NLL is roughly $1 : -9 : -4$. We
see (as observed already in Ref.~\cite{Brod:2013cka}) that the
contribution of the photonic Barr--Zee diagram is negligible. We also
see that the NLL correction is quite large, about half the size of the
LL contribution.

The leading terms in the expansion of the solution for the RG equations
for the chromoelectric dipole, on the other hand, should exactly reproduce the
logarithmic parts of the result in Eq.~\eqref{eq:dsw}
(the constant term is of next-to-next-to-leading logarithmic order and
can only be reproduced, within EFT, by a three-loop calculation).
Indeed, we find
\begin{equation}
\begin{split}
	\tilde d_q  =& - \sqrt{2} G_F \frac{m_q}{g_s^2} C_4^q (m_b) \\
	       \simeq& - \sqrt{2} G_F \frac{m_q}{g_s^2} \bigg(
  \frac{\alpha_s(m_b)}{4\pi} \bigg)^2 \bigg\{
	  C_1^{\q\qp(0)} (M_h)    \frac{\gamma_{1,\q\qp \to 4,\q\qp}^{(0)} \gamma_{4,\q\qp \to 4,\q}^{(0)}}{8} \log^2 x_b \\
	&\hspace{10.2em}-C_1^{\q\qp(0)} (M_h) \frac{\gamma_{1,\q\qp \to 4,\q}^{(1)}}{2} \log x_b\\
	&\hspace{10.2em}-C_4^{\q\qp(1)} (M_h) \frac{\gamma_{4,\q\qp \to 4,\q}^{(0)}}{2} \log x_b\bigg\} \\
	=& 2 \frac{\alpha_s(m_b)}{(4\pi)^3}
  \sqrt{2} G_F m_q \kappa_b \sin\phi_b x_b \log^2 x_b + 0 \,.
\end{split}
\end{equation}
As expected, the leading contribution to the LL reproduces the
quadratic logarithm in Eq.~\eqref{eq:dsw}, while leading contribution
to the NLL result vanishes.

This means, in turn, that the NLL corrections to $\tilde d_q$ start at
relative order $\alpha_s/(4\pi)$, with large anomalous-dimension
prefactors of $(\gamma_{1,\q\qp \to 1,\q\qp}^{(0)} \gamma_{1,\q\qp \to
  4,\q}^{(1)} + \gamma_{1,\q\qp \to 4,\q\qp}^{(0)} \gamma_{4,\q\qp \to
  4,\q}^{(1)})/8 = 47$ and\\$(\gamma_{1,\q\qp \to 4,\q}^{(1)}
\gamma_{4,\q \to 4,\q}^{(0)} + \gamma_{1,\q\qp \to 4,\q\qp}^{(1)}
\gamma_{4,\q\qp \to 4,\q}^{(0)})/8 = 404/9$, multiplied by the LO
initial condition $-1$, and $(\gamma_{4,\q\qp \to 4,\q}^{(0)}
\gamma_{4,\q \to 4,\q}^{(0)} + \gamma_{4,\q\qp \to 4,\q\qp}^{(0)}
\gamma_{4,\q\qp \to 4,\q}^{(0)})/8 = 56/3$, multiplied by the NLO
initial condition $3/2$. Therefore, these constitute sizeable relative
corrections, with ratio $1 : -0.6$ between the LL and the NLL
contributions. Needless to say that in our numerics we use the full
solution of the RG equations, where all large logarithms are resummed
to leading and next-to-leading order.

For completeness we give the also expansion of the Dicus
result~\cite{Dicus:1989va}:
\begin{equation}
\begin{split}
	w  = & - \frac{\sqrt{2} G_F}{g_s} C_w (m_b) \\
	       \simeq& - \frac{\sqrt{2} G_F}{g_s} \bigg(
  \frac{\alpha_s(m_b)}{4\pi} \bigg)^2 \bigg\{
	  - \frac{1}{2} C_1^{\q(0)} (M_h) \frac{\gamma_{1,\q \to 4,\q}^{(0)}}{2} \log x_b \bigg\} \\
	=& - \sqrt{2} G_F \frac{g_s(m_b) \alpha_s(m_b)}{(4\pi)^3}
           \kappa_b^2 \sin\phi_b \, \cos\phi_b \, x_b \log x_b \,.
\end{split}
\end{equation}
This reproduces the logarithmic term in Eq.~\eqref{eq:dsw}.

\addcontentsline{toc}{section}{References}
\bibliographystyle{JHEP}
\bibliography{references}

\end{document}